\documentclass[structabstract]{aa}  
\usepackage{epsfig}
\usepackage{graphicx}
\usepackage[varg]{txfonts}
\usepackage{color}
\usepackage{xspace}
\usepackage[dvipsnames]{xcolor}

\newcommand{\rnoff}{\color{black}\xspace}

\usepackage{natbib}
\usepackage{longtable,lscape}
\usepackage{rotating}
\usepackage[normalem]{ulem}

\usepackage{xcolor,cancel}

\newcommand{\BP}{$G_{\rm BP}$}
\newcommand{\RP}{$G_{\rm RP}$}

\newcommand{\Gaia}{\emph{Gaia}}

\newcommand{\GDR}{\Gaia~DR}
\newcommand{\GEDR}{\Gaia~EDR}

\usepackage{verbatim}
\usepackage{color}
\usepackage{xcolor}

\bibpunct{(}{)}{;}{a}{}{,} 

\newcommand{\orcit}[1]{\protect\href{https://orcid.org/#1}{\protect\includegraphics[width=8pt]{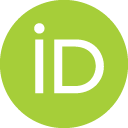}}}

\usepackage{hyperref}
\hypersetup{
    colorlinks=true,
    linkcolor=blue,
    filecolor=magenta,      
    urlcolor=cyan,
}

\begin{document}

   \title{Internal calibration of {\Gaia} BP/RP low-resolution 
spectra 
}
\titlerunning{Internal calibration of {\Gaia} BP/RP low-resolution 
spectra}
\authorrunning{J.M. Carrasco et al.}


   \author{
J.M.      ~Carrasco                      \orcit{0000-0002-3029-5853}\inst{\ref{inst:0013}}
\and M.        ~Weiler                        \orcit{0000-0002-3007-3927}\inst{\ref{inst:0013}}
\and C.        ~Jordi                         \orcit{0000-0001-5495-9602}\inst{\ref{inst:0013}}
\and C.        ~Fabricius                     \orcit{0000-0003-2639-1372}\inst{\ref{inst:0013}}
\and F.        ~De Angeli                     \orcit{0000-0003-1879-0488}\inst{\ref{inst:0009}}
\and D.W.      ~Evans                         \orcit{0000-0002-6685-5998}\inst{\ref{inst:0009}}
\and F.        ~van Leeuwen                   \inst{\ref{inst:0009}}
\and M.        ~Riello                        \orcit{0000-0002-3134-0935}\inst{\ref{inst:0009}}
\and P.        ~Montegriffo                   \inst{\ref{inst:0036}}
          }

\offprints{J.M.~Carrasco (carrasco@fqa.ub.edu)}

   \institute{
Institut de Ci\`{e}ncies del Cosmos (ICCUB), Universitat  de  Barcelona  (IEEC-UB), Mart\'{i} i  Franqu\`{e}s  1, 08028 Barcelona, Spain\relax                                                                                                                                                              \label{inst:0013}
\and Institute of Astronomy, University of Cambridge, Madingley Road, Cambridge CB3 0HA, United Kingdom\relax                                                                                                                                                                                                    \label{inst:0009}
\and INAF - Osservatorio di Astrofisica e Scienza dello Spazio di Bologna, via Piero Gobetti 93/3, 40129 Bologna, Italy\relax                                                                                                                                                                                    \label{inst:0036}
             }

   \date{\today, Received / Accepted}


  \abstract
    {
    The full third {\Gaia} data release will provide, for the first time, the calibrated spectra obtained with 
the blue and red {\Gaia} slitless spectrophotometers (BP and RP, respectively).
{\Gaia} is a very complex mission and cannot be considered as a single instrument, but rather as many instruments. The two lines of sight with wide fields of view (FoV) introduce strong variations of the observations across the large focal plane with more than one hundred different detectors.
The main challenge when facing {\Gaia} spectral calibration is that no lamp spectra or flat fields are available during the mission. 
Also, the significant size of the line spread function (LSF) with respect to the dispersion of the prisms produces alien photons contaminating neighbouring positions of the spectra.
This makes the calibration special and different from standard approaches. 
    }
   {
   This work gives a detailed description of the internal calibration model
   for the spectrophotometric data used to obtain the content of the {\Gaia} catalogue.   
   The main purpose of the internal calibration is to bring all the epoch spectra onto a common flux and pixel (pseudo-wavelength) scale, taking into account variations over 
the focal plane and with time, producing a mean spectrum from all the observations of the same source.
}
   {
In order to describe all observations on a
common mean flux and pseudo-wavelength 
scale, we constructed a suitable representation of the internally calibrated mean spectra via basis functions, and we described the transformation between non-calibrated epoch spectra and calibrated mean spectra via a discrete convolution, 
parametrising the convolution kernel to recover the relevant coefficients.
   }
   {
The model proposed here for the internal calibration of the {\Gaia} spectrophotometric observations 
is able to combine all observations into a mean instrument to allow the comparison of different sources and observations obtained with different instrumental conditions along the mission and the generation of mean spectra from a number of observations of the same source. 
We derived a calibration model that can handle the self-calibrating nature of the problem.
The output of this model provides the internal mean spectra, not 
as a sampled function (flux and wavelength), but as a linear combination of basis functions, although sampled spectra can easily be derived from them.
}
   {}
   \keywords{Instrumentation: spectrographs; Space vehicles: instruments; Techniques: spectroscopic
; Galaxy: general; Stars: general}
   
  \maketitle
  
%

\rnoff

\section{Introduction}
\label{sec:introduction}

The {\Gaia} space mission \citep{Prusti2016}
launched by the European Space Agency (ESA) in 2013, 
is producing a full sky survey for all sources brighter than 20th magnitude. 
The {\Gaia} Early Data Release 3 ({\GEDR3}, see \citealt{Brown2021}) includes 1.8 billion sources, 
representing about 1\% of the content of the Galaxy. 
Besides the astrometric and kinematic information (position, parallaxes, proper motions, and radial velocities), the photometric and spectrophotometric data are also one of the main products of 
the {\Gaia} mission, which allow us to derive the astrometric chromaticity correction and to determine the astrophysical information of the observed sources. The first spectrophotometric mean spectra will 
be published with the full {\Gaia} Data Release 3 ({\GDR3}), which is planned for 2022\footnote{Visit the {\Gaia} data release information web page (\href{https://www.cosmos.esa.int/web/gaia/release}{https://www.cosmos.esa.int/web/gaia/release}) for updated dates and content of every {\Gaia} data release.}.

During the mission, no flat fields nor spectroscopic lamp images were obtained at any point. Because of the large number of instrumental configurations and the accuracy level we want to reach, the information to perform the calibration of the instrument needs to be derived from the scientific data itself, as a small set of calibrators would not provide enough information in relation to the complexity of the instrument. Thus, the road of self-calibration is essentially the only option for the construction of the internal reference system.

\begin{figure}[!htbp]
   \centering
\includegraphics[width=0.9\columnwidth]{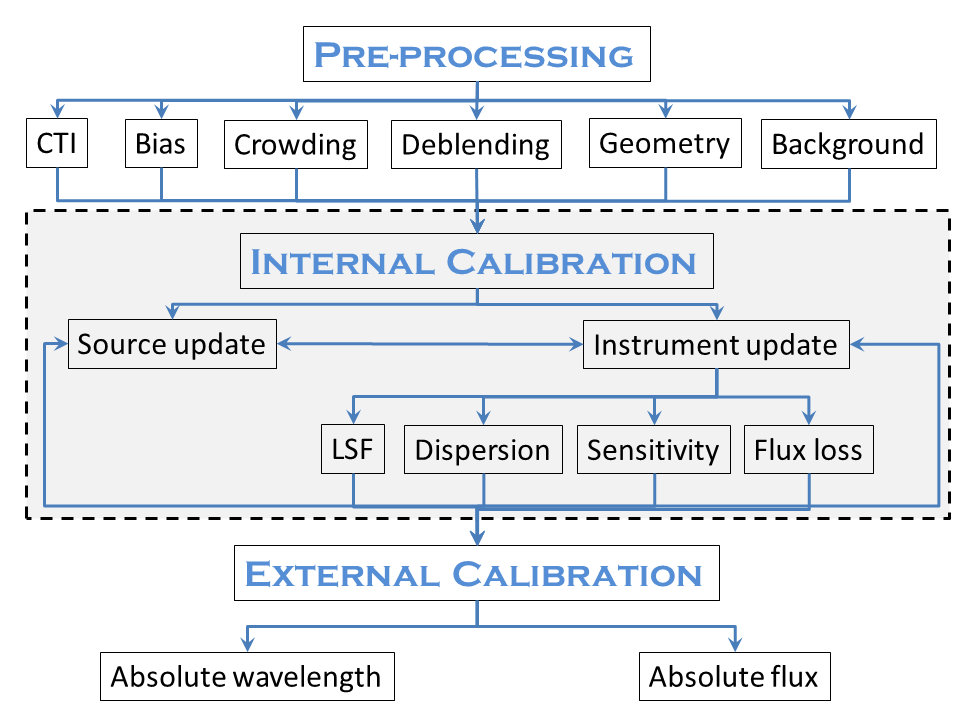}
\caption{Scheme used to calibrate {\Gaia} spectrophotometry. This paper only deals with the internal calibration piece (dashed square). 
}
  \label{fig:scheme}
\end{figure}

The spectrophotometric calibration is complex, with important instrumental effects to account for (see Sect.~\ref{sec:effects}). For instance, the dispersion of the spectra and the point spread functions (PSF) change significantly across the large {\Gaia} focal plane. Because of the complexity of the problem, 
the calibration of these spectra is separated conceptually and 
organisationally into three distinct parts (as is also the case for the integrated photometry calibration; \citealt{Carrasco2016,Riello2021}). 
As can be seen in Fig.~\ref{fig:scheme}, after an initial preparation step called pre-processing, the internal (or relative) calibration uses only {\Gaia} observations to relate all observations into a common system. 
After this internal calibration, the external (or absolute) calibration uses a relatively small number of ground-based standard stars \citep{pancino2012,pancino2021,altavilla2021}
to get the absolute fluxes and wavelengths. 
This paper describes the 
concepts of the internal calibration of {\Gaia} 
spectrophotometric data, but this is not linked to the particular details used in any of the {\Gaia} data releases. Different papers will be published, together with {\GDR3} and future releases, with the particular details of the actual model used to produce that particular dataset (i.e. giving the actual used basis functions for the source representation, the set of calibrators, the instrument model parameters, etc.). Pre-processing and external calibration procedures also fall outside the scope of this paper. 

The internal calibration (Sect.~\ref{sec:basicformulations}) models the instrument without the use of external information,
working entirely in the instrument system.
The way to proceed without external information is to study how a set of (assumed constant) calibration sources (Sect.~\ref{sec:calibrators}) produce different 
observations when obtained with different instrumental conditions: how they change with time, the focal plane position, detector used, field of view (FoV), and so on. Section~\ref{sec:effects} explains these instrumental effects and how the key point when performing the internal calibration is to model the relative differences among observations (in terms of dispersion, line spread function, sensitivity, etc.), not their absolute values.
As neither the instrument nor the sources are known in advance, this process needs to determine both iteratively (Sect.~\ref{sec:instrumentcalibration}). Sections~\ref{sec:basicformulations} and \ref{sec:source}, respectively, describe the models representing the instrument and the sources, defining some parameters to be determined during the calibration process. Section~\ref{sec:results} 
shows preliminary results of the internal calibration model explained here.
Finally, Sect.~\ref{sec:conclusions} includes a summary and conclusions of this work.

\section{The {\Gaia} spectrophotometers}
\label{sec:effects}

In this section, we briefly review the {\Gaia} spectrophotometers and the instrumental effects 
affecting the observations to be calibrated, 
and we refer the reader to several previous papers where more detailed descriptions of these effects can be found (\citealt{Prusti2016,Brown2018,Riello2021,Hambly2018,Crowley2016,Fabricius2016} and \citealt{Carrasco2016}). Here, we are only interested in describing the concepts needed in our modelling (see Sect.~\ref{sec:instrumentcalibration}).

\begin{figure*}[t!]
 \centering
 \includegraphics[width=0.9\linewidth]{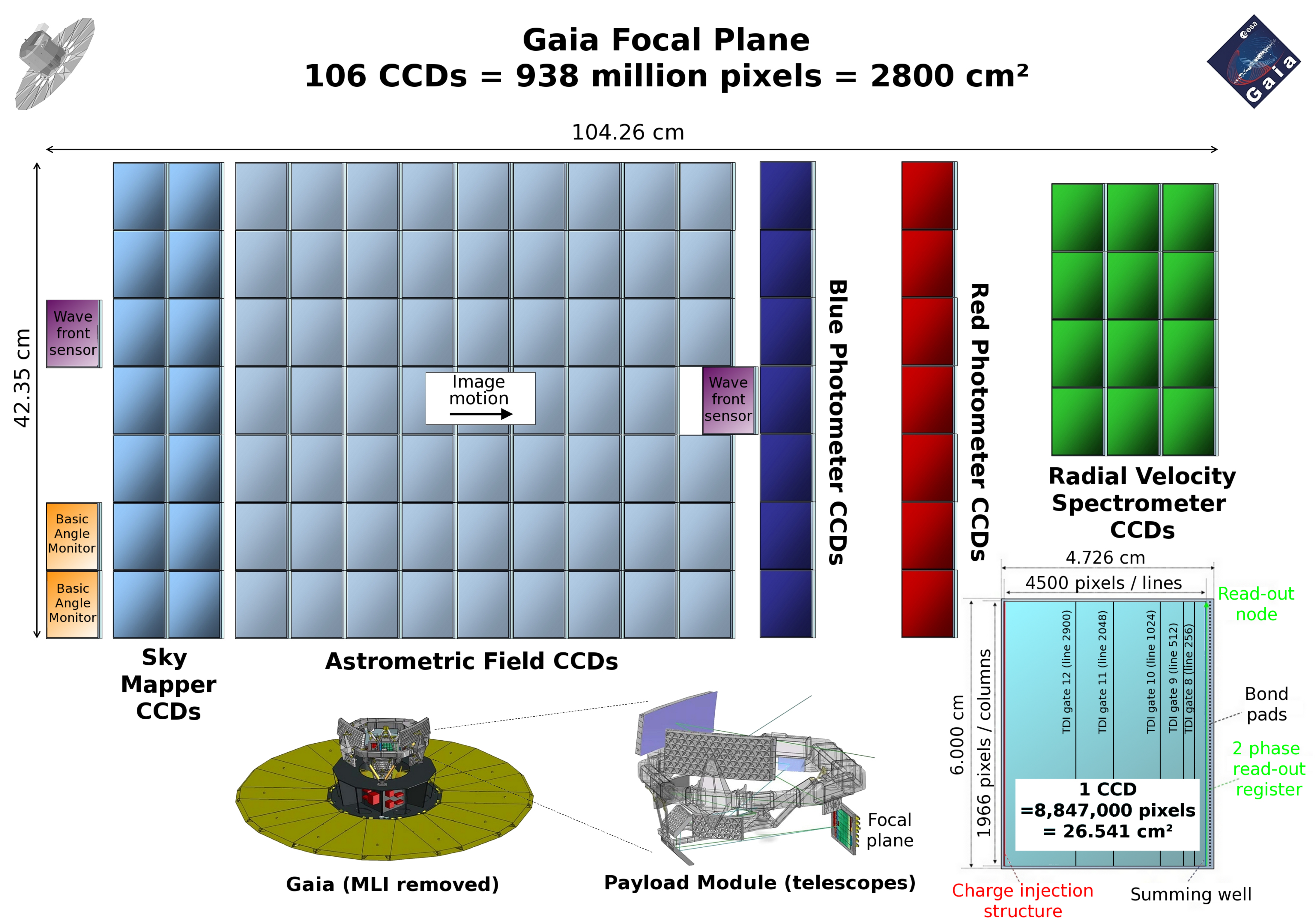}
  \caption{ {\Gaia} focal plane. For the photometric calibration explained here, only CCDs in the blue and red photometers are relevant (picture courtesy of ESA - A. Short).
 \label{fig:focalplane}
}
\end{figure*}

The {\Gaia} satellite collects the light from two different telescopes (FoVs). As {\Gaia} is spinning while observing the sky, in one of these revolutions one of these telescopes (the leading FoV) is always first to observe a region in the sky, while the second one will observe it afterwards (the following FoV). The light from these two different telescopes is projected onto a common focal plane with charge-coupled device (CCD) detectors. 
For the case of the {\Gaia} spectrophotometers, 
after being dispersed by two different 
slitless prisms, the light is collected by one set of CCDs for each spectrophotometer (with seven CCD rows\footnote{We use CCDs 1 to 7 to refer to the different CCD rows in the spectrophotometric instrument.} across the focal plane, see Fig.~\ref{fig:focalplane}). 
The spectral range covered by the two spectrophotometers 
is different. The blue photometer (BP) covers the short wavelengths in the 330--680~nm range, while the red photometer (RP) covers longer wavelengths, in the 640--1050~nm range. 

In order to reduce the telemetry of the satellite, only a small window around the detected position of the source is downloaded. 
The size of the window ($60$~pixels~$\times\ 12$~pixels~$=3.5$~arcsec~$\times\ 2.1$~arcsec) around the spectra was selected 
to be wide enough to include most of the flux from the source 
once dispersed by the spectrophotometric prisms and also contain a range of pixels for background flux determination. The longer size of the BP and RP windows (60 pixels, equivalent to $3.5$ arcsec) in comparison with the window scheme used for the astrometric field observations (only 12 or 18 pixels wide, equivalent to $0.7$ and $1.1$~arcsec, respectively) implies a higher chance of overlapping windows (triggering truncation) and blends (where more than one source is captured in the same window).

As previously mentioned, {\Gaia} is spinning on itself in order to scan all of the sky. 
This spinning process causes the images of the sources in the sky to transit the focal plane during the observation. 
We refer to the transit direction as 'along scan' (AL).
The perpendicular direction to AL is called 'across scan' (AC).
Each CCD is operated in time delay integration (TDI) mode to compensate for this spinning effect.
TDI mode consists of reading the image not at the end of a predefined time exposure, but continuously at a given rate. 
The charges are clocked through the CCD in the AL direction at a constant rate, and the satellite rotation is constantly adjusted to make the images move at the same rate
as the reading takes place. With this procedure, the light from every source is integrated during the transit over every CCD.

The result for {\Gaia} is that the observed source transits each CCD in about 4.4 seconds 
(this being the maximum exposure time of each observation). 
The use of 'gates' limits the region of the CCD to be read, effectively limiting the exposure time in case the source is too bright.
Each of the CCDs has 4500 pixels in the AL direction and 1966 pixels in the AC direction.
The recording process of the BP and RP spectrophotometry with the CCD devices produces a pixelisation of the spectra that is highly 
dependent on the geometry of the CCD and its pixels, the allocation of the reading window around the observation, and so on.

The 2D image of the spectra in BP and RP is collapsed into 1D spectra, binning the pixels in the AC direction while being read for observations fainter than $G=11.5$~mag.
This introduces the concept of a 'sample', 
corresponding to a group of 12 pixels in the AC direction once combined to constitute the 1D spectra. In this way, these 1D spectra are composed of 60 different samples.
Nevertheless, the outermost ten samples (approximately) on either side of the window correspond to a negligible instrumental response, and the signal there is dominated by background flux and the smearing effects of the line spread function (LSF).

The calibration of the spectrophotometric observations is complicated due to intrinsic variations 
of the instrumental conditions for different observations. The {\Gaia} focal plane is very large\footnote{The full {\Gaia} focal plane is 104~cm $\times$ 42~cm. BP and RP CCDs also cover a length of 42 cm, but they are limited to the size of one single CCD strip (of about 4.5 cm wide).}, with a large number of CCDs\footnote{The full {\Gaia} focal plane comprises 106 different CCDs. Among them, each BP and RP instrument uses seven different CCDs.}, two overlapping FoVs (with their different optical paths), the occurrence of gates and window strategies, and the instrumental behaviour changing with time. The main instrumental ingredients to be considered for the calibration of the {\Gaia} spectrophotometry are geometry, dispersion, LSF, sensitivity and flux loss. We explain them in detail in the following paragraphs.

The geometric calibration deals with the fact that different spectra obtained at different positions of the focal plane are not perfectly aligned. We need to know the differences in the actual projection of the spectra on the detectors.
The dispersion direction may also not be perfectly aligned with the transit direction, and thus some tilt could be present. This will produce a discrepancy between the expected position in the CCD of a given wavelength and the actual one. 
The instrumental model explained in this work can cope with small spectral misalignments in the AL direction (limited by the maximum number of neighbours 
considered in the model, see Sect.\ref{sec:basicformulations}), but larger misalignments need to be corrected by pre-processing before our model is applied.

\begin{figure}[!htbp]
   \centering
\includegraphics[width=0.9\columnwidth]{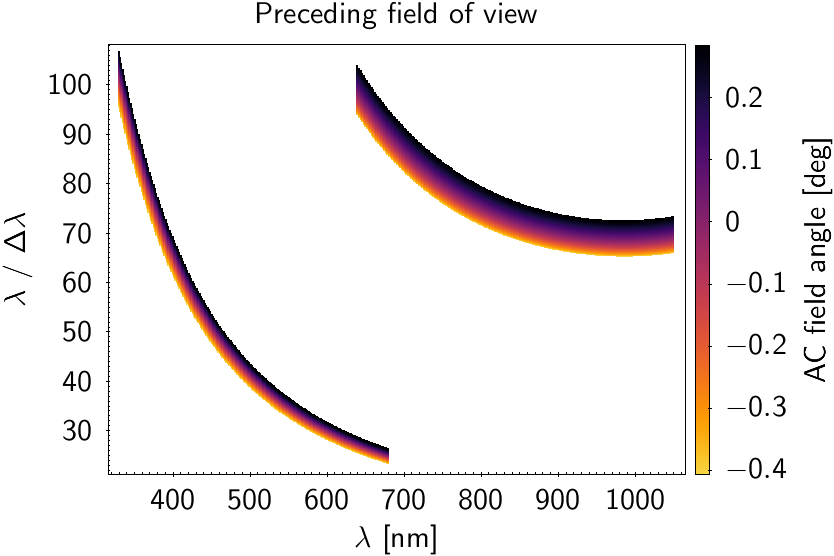}\\
\hspace{-1.1cm}
\includegraphics[width=0.76\columnwidth]{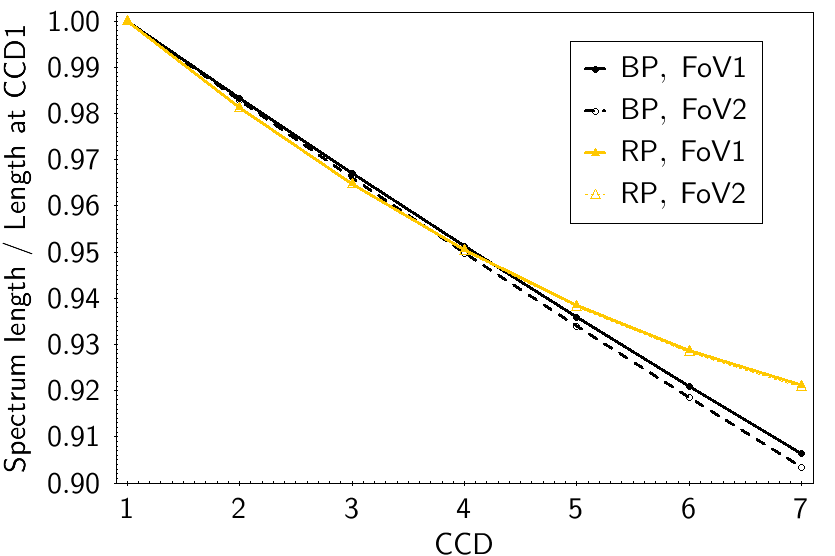}
\caption{{\it Top:} Nominal spectral resolution relationship\protect\footnotemark\
 with wavelength ($\lambda$). $\Delta\lambda$ is the width of a spectrophotometric sample for BP (shorter wavelengths) and RP (longer wavelengths) spectrophotometers for the preceding telescope. The colour scale indicates how this resolution changes with the AC position in the focal plane. {\it Bottom:} Length of the BP (black lines) and RP (orange lines) spectra for different CCDs and FoVs compared with the length in CCD1.}
  \label{fig:resolution}
\end{figure}

The spectral dispersion could also be different for different observations. 
As the initial purpose of the two spectrophotometers is mainly to provide the chromatic centroid correction in astrometry \citep{Lindegren2018,Lindegren2021}
as a function of the spectral energy distribution of the source, the spectral resolution of these two instruments is lower than 100 
and depends on the wavelength range and the position in the focal plane (see Fig.~\ref{fig:resolution}, top). Although similar, the dispersion law for the two different FoVs is also different. The variation of the dispersion law with the position in the focal plane also causes the length of the spectra to be different on different CCDs (see Fig.~\ref{fig:resolution}, bottom).

The resolution of the instrument was chosen to be able to reproduce the passbands described in \cite{Jordi2006}. 
In spite of this low resolution, the amount of science potentially 
available for so many sources in all the sky with homogeneous spectrophotometric observations is very 
promising. This is noted, for example, in Sect.~\ref{sec:results}, where some calibrated spectra for sources with different colours are shown, making it possible to distinguish their different spectral energy distributions. The astrophysical parameters derived from the calibrated spectrophotometry will also be published with the next {\GDR3} catalogue.

Besides the variation of the dispersion properties at different positions across the focal plane, the wavelength range at different positions can be slightly different because of the passband differences, providing different widths for the spectra. Table~\ref{tab:bandwidth} shows the cut-on and cut-off wavelengths ($\lambda_{\rm cuton}$ and $\lambda_{\rm cutoff}$ respectively), which translates to different total bandwidths in 
wavelengths ($\Delta =\lambda_{\rm cutoff}-\lambda_{\rm cuton}$).
Due to these effects, the correspondence between samples and wavelengths will vary for different observations of a given source.  
Furthermore, the spectral dispersion is also expected to change with time. 
All these dispersion variations (summed to the more important effect of the AL centring of the windows previously explained) prevent us from naively summing the different epoch spectra to obtain a mean spectrum.

\footnotetext{\url{https://www.cosmos.esa.int/web/gaia/resolution}}

\begin{table}
\begin{center}
\caption{Bandwidth variations in different {\Gaia} CCDs.
\label{tab:bandwidth}}
\begin{tabular}{cccc}
\hline
                            & BP           & RP           \\
\hline
$\lambda_{\rm cuton}$ [nm]  & $397$--$399$   & $629$--$631$ \\ 
$\lambda_{\rm cutoff}$ [nm] & $665$--$668$   & $918$--$919$ \\ 
$\Delta$ [nm]               & $266$--$270$   & $287$--$290$ \\
\hline          
\end{tabular}
\end{center}
\end{table}

\begin{figure}[!tbp]
   \centering
\includegraphics[width=0.9\columnwidth]{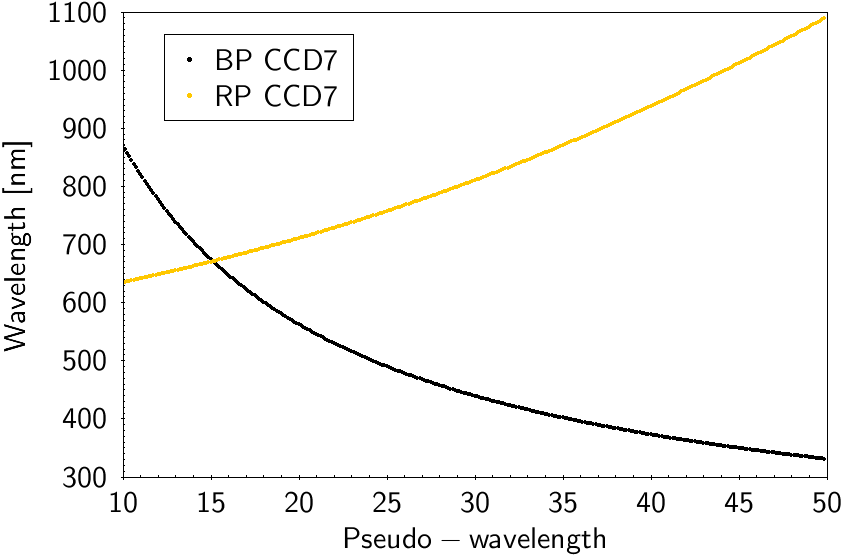}\\
\caption{Correspondence between the absolute wavelength and pseudo-wavelength scale for simulations in BP (black) and RP (orange) used in this paper for CCD7. These are only intended to be used as indicative values for the reader.}
  \label{fig:wavelength}
\end{figure}

Once the geometric calibration and a rough dispersion law is known, we can define an internal reference scale, referred to as 'pseudo-wavelength'. 
We call it pseudo-wavelength because it does not yet have the usual physical units for wavelengths (nanometres, for instance), 
but it is measured in units of sample in an arbitrarily chosen reference location in the focal plane (AL positions in the 1D window for a given CCD and AC coordinate). 
The external calibration will produce the transformation to the physical units.
As a reference, pseudo-wavelength values 10 and 50 approximately correspond to wavelengths 860 and 330 nm in BP and 630 and 1090 nm in RP (see Fig.~\ref{fig:wavelength}).
In the internal calibration process explained in this work, this 
pseudo-wavelength scale is sufficient in order to align the observed spectra and form the mean spectra from various observations of a given source obtained at different locations of the focal plane, FoV and time. 
The external calibration step (see Fig.~\ref{fig:scheme}) provides the relation between pseudo-wavelengths and true wavelengths.

\begin{figure}[!htbp]
   \centering
\includegraphics[width=0.8\columnwidth]{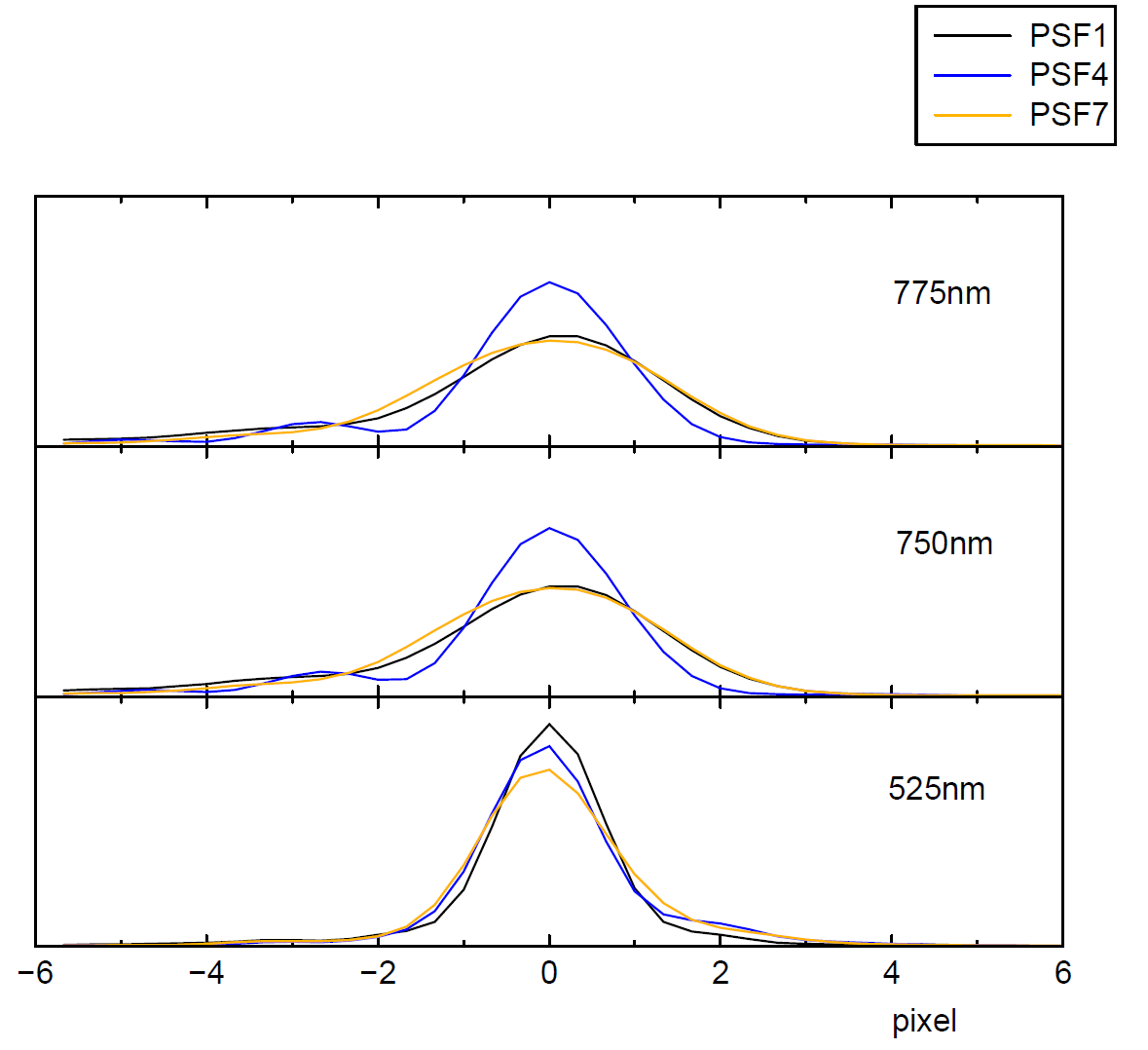}
\caption{LSF at three different wavelengths (525~nm at MgH band in the bottom panel, 750~nm at pseudo-continuum region in the central panel, and 775 nm at TiO band in the top panel). In each panel, three different CCDs are plotted, corresponding to the bottom (CCD1 in black), middle (CCD4 in blue), and upper (CCD7 in orange) positions of the focal plane, respectively.
  }
  \label{fig:lsf}
\end{figure}

Due to diffraction, aberrations, and further non-optical effects, such as the CCD pixelisation or spacecraft motion, the image of a point-like source obtained with a telescope  
is not a point, but it has some spread on the focal plane. The final image of this point obtained by the detector is known as PSF. The LSF can be understood as analogous to the PSF when collapsed 
into 1D.
When this concept is applied to spectroscopic images, the result is that the light with a 
given wavelength is not registered only in a single pixel, as desired, but in several, 
depending on the width of the LSF (1-2 pixels wide, see Fig.~\ref{fig:lsf}). This produces some contamination of the adjacent pixels by light with wavelengths 
nominally corresponding to the neighbouring pixels according to the dispersion law.
This effect (sometimes referred to as the alien photon problem) changes with wavelength, the position in the focal plane (see Fig.~\ref{fig:lsf}), the FoV, and time. 

\begin{figure*}[!htbp]
   \centering
\includegraphics[width=0.32\linewidth]{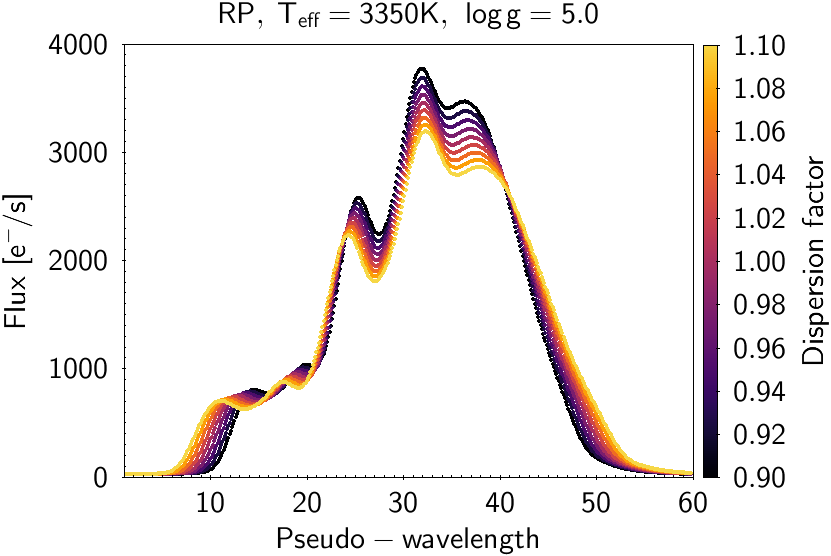}
\includegraphics[width=0.32\linewidth]{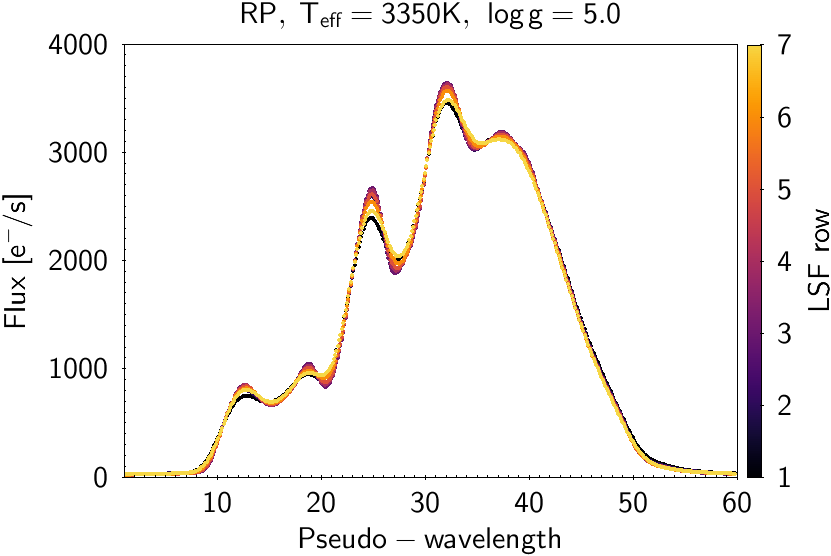}
\includegraphics[width=0.32\linewidth]{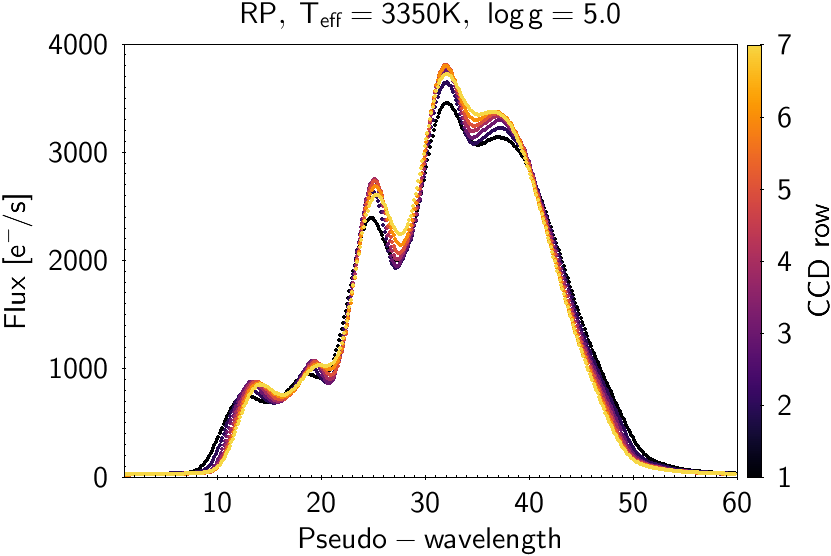}
\caption{RP simulations, using the synthetic spectral energy distribution library of \cite{Lejeune1997}, of the effect of $\pm 10\%$ dispersion variation (left), 
LSF conditions in the seven CCDs (centre), and both effects together as expected in each CCD (right), 
for a star with an effective temperature 
of $T_{\rm eff}=3350$~K.
The $\pm 10\%$ dispersion variation considered in the left plot represents the maximum variation expected for the different observations. The dispersion at a given time of the mission changes across the focal plane with the CCD. The latter effect is included in the plot at the right.
  }
  \label{fig:instrumeffect}
\end{figure*}

Therefore, the smearing of the spectrum will be different for each observation. 
Figure~\ref{fig:instrumeffect} illustrates this, showing first the dispersion and LSF effect alone 
(left and central panels) and then combining them in the right panel. The simulations in Fig.~\ref{fig:instrumeffect} were produced using the {\Gaia} Object Generator (GOG, \citealt{luri2014}). The evolution of the contamination of the mirrors, refocussing events and changes due to environmental conditions (such as the effects of the decontamination events) affect the LSF (see \citealt{Riello2021,Rowell2021} and \citealt{Mora2014}).

Flat field images cannot be obtained during the {\Gaia} mission  
to measure sensitivity variations, as is commonly done with on-ground instruments. 
The only flat field information available was obtained before launch for the sole purpose of obtaining an estimation of the instrument's behaviour. It is not suitable to describe the sensitivity variations during the mission. The different episodes of contamination on the mirrors and the detectors, the decontamination events, and the expected ageing of the optics and CCDs (see \citealt{Riello2021}) make these pre-launch estimations insufficient for covering the mission. 
The response of the instrument to a given wavelength will be different for different FoVs, 
CCDs, gates, and positions on the CCD. As the observation is integrated in the AL direction, only the AC position (column) in the CCD is relevant here for the sensitivity analysis. 
Within each CCD, a column response non-uniformity (CRNU) of a 
few per cent is present \citep{Crowley2016}.

\begin{figure*}[!htbp]
   \centering
\includegraphics[width=0.495\textwidth]{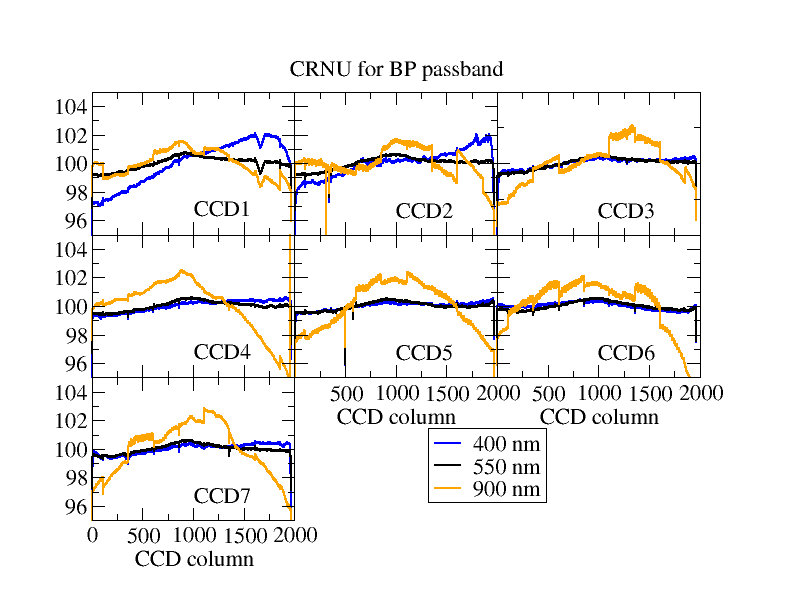}
\includegraphics[width=0.495\textwidth]{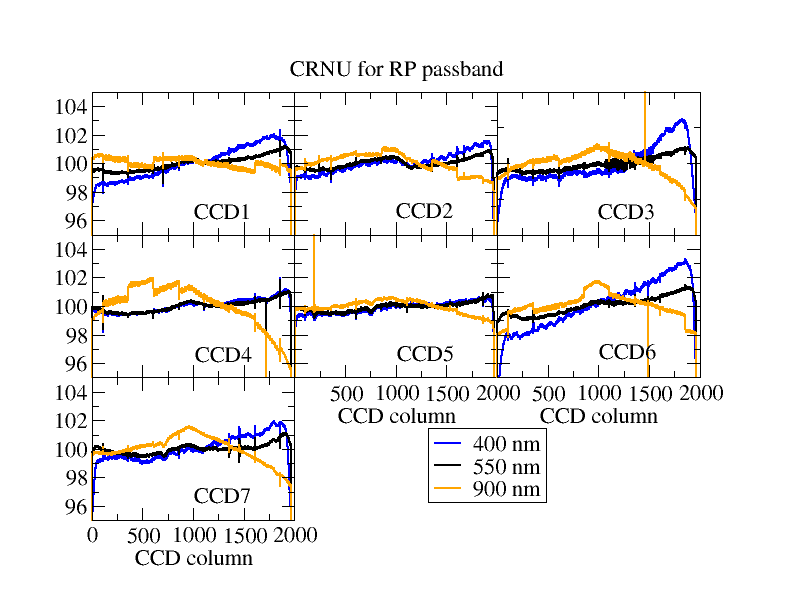}
\caption{Column response non-uniformity (CRNU; \citealt{Kohley2009}). Dependence of the response of the CCD detector as a function of the AC position (column) and CCD for BP (left) and RP (right) for three different wavelengths (in different colours). Some abrupt changes in neighbouring AC positions can be observed in the different CCDs due to stitch block discontinuities.
  }
  \label{fig:flatfield}
\end{figure*}

Figure~\ref{fig:flatfield} shows some examples of these CRNU images obtained before launch. 
The lower pixel responses are found at the corners of the CCD. The sensitivity variation is mostly smooth with the column, 
except at the stitch boundaries inside the CCD. The stitch boundaries are the result of the fact that a given CCD 
is produced in several photo-lithographic units. The separation between these parts produces small jumps in the sensitivity of the device in neighbouring columns. Moreover, dead pixels or hot ones can 
also produce variations in the flux obtained at discrete positions.
A degradation of the sensitivity with time is also to be expected.

The flux loss is the combination of the LSF effect with the limited extension of the window around the source 
being transferred to ground. It results in part of the light of the spectra being lost in our measurements. The imperfect 
centring of the window around the source also increases the amount of flux loss.

By simulating observations with nominal LSFs we can estimate the influence of the flux loss in {\Gaia} spectrophotometry. The total amount of flux lost outside the window is about $4$--$5\%$ in the case of BP and $4.5$--$5.5\%$ in the case of RP. 
The reason for having more flux outside RP windows is that the LSF is wider for larger wavelengths. 
Added to this, the centring error in the RP observations is slightly larger due to the RP zooming effect (see \citealt{Riello2021}), producing a larger flux loss than in BP.

The relevant parameter for the internal calibration is the variation of the flux loss in different observations. The flux loss can change by about $0.8\%$--$1.4\%$ when shifting the window $\pm 2$ pixels in the AC direction 
and considering different AC motion due to the intrinsic motion of the astrophysical source and the satellite attitude. 
The effect of the AC shift does not depend much on the colour of the source and the variations are smaller 
than $0.05\%$, meaning that the variation is almost grey.

In the case of the AL flux loss, 
the variation in flux is limited to $0.5\%$
when a $\pm 3$ pixel shift in the AL direction is considered.
This AL flux loss can change with colour ($0.2\%$--$0.6\%$ in BP and $0.2\%$ in RP). 
The maximum variation between CCDs of the flux inside the window 
is about $\pm 0.02$--$0.03\%$ in the case of an AL shift$=+1$ (i.e. very little compared with changes due to AC shifts).
These estimations do not account for the extra losses due to bad focus periods during operations.

Observations in different FoVs meet partially separate optical surfaces and have different optical paths, which may result in different dispersed LSFs and response functions. Gated integration over a different section of the CCD for bright sources implies different dispersed LSF and response conditions. To take this complexity in the {\Gaia} spectrophotometers into account, we propose the calibration model explained in the following sections.

The presence of blended observations, non-perfect bias and background subtraction, charge injection events, dead pixels, 
and other non-linear effects such as saturation and charge distortion complicates the calibration process even more 
(see \citealt{Riello2021,Evans2018,Hambly2018,Crowley2016} and \citealt{Fabricius2016} for more details). 
The calibration pipeline must account for as many of these effects as possible, or at least try to evaluate 
the extra uncertainty in the final result if they are not considered.
Another (or complementary) approach is to simply reject some of the observations affected by these unaccounted effects as outliers, 
if they represent a small part of the observations for that source. The remaining observations of a given source can be used to 
build a mean spectrum, which will reduce the uncertainty in principle by a factor equal to the square root of the number of observations considered.

The number of observations for a given source depends on the satellite scanning law, and the probability of detecting a source 
depends on its magnitude and the crowding in the observed region of the sky (also considering the second FoV, 
which overlaps in the same focal plane). On average, a source in the {\Gaia} catalogue will have about one hundred observations in five years. The very large volume of data to be processed adds to the challenge. 
For reference, by the time of {\GEDR3} (December 2020), more than 300 billion spectrophotometric observations had been obtained\footnote{We refer the reader to an updated estimation for the number of observations in each {\Gaia} instrument at \href{https://www.cosmos.esa.int/web/gaia/mission-numbers}{https://www.cosmos.esa.int/web/gaia/mission-numbers}.}, all of them under different instrumental conditions. Given the size and complexity of the task, a priority requirement is to keep 
the load on the computer systems at a feasible level using a robust pipeline.

\section{Basic formulations}
\label{sec:basicformulations}

\begin{figure*}[!htbp]
   \centering
\includegraphics[width=0.8\textwidth]{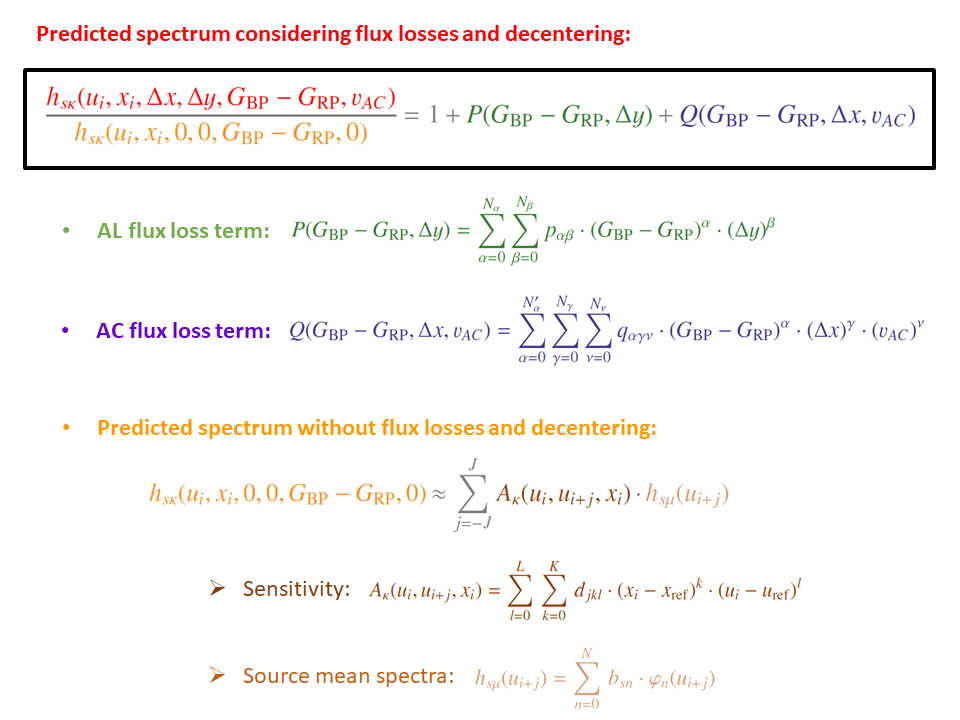}
\caption{Summary of the equations explained in Sects.~\ref{sec:basicformulations} and \ref{sec:instrumentcalibration}. The meaning of the variables is detailed in the text and in Table~\ref{tab:variables}.
  }
  \label{fig:SummaryEquations}
\end{figure*}

In this section and the following one we establish the main equations of the model, which are summarised in Fig.~\ref{fig:SummaryEquations}. BP and RP spectra of the {\Gaia} mission are internally calibrated 
independently of each other. The calibration approach is exactly the same, however. 
In order to be concise, we do not refer to BP and RP spectra or BP- and RP-specific indices. 
Instead, we simply speak of 'spectra', which can mean either the BP spectra or RP spectra.\par

The different instrumental properties (Sect.~\ref{sec:effects}) drive the obtained epoch spectrum of a given source. We define $\omega_1,\ldots,\omega_\Omega$ as the set of $\Omega$ parameters needed to describe the instrument influence on the epoch observation. The parameters $\omega_1,\ldots,\omega_\Omega$ required to describe the differences among instruments and, consequently, the epoch spectra, can either be continuous (for example, time variations and AC position variations) or discontinuous or discrete (FoV, gates, window class, decontamination and refocusing events, CCD stitch blocks, etc.).

We may write the epoch spectrum for a source, $s$, in a given pseudo-wavelength, $u$, as $h_s(u,\omega_1,\ldots,\omega_\Omega)$. 
We furthermore assume that each epoch spectrum is linked to the spectral photon distribution (SPD, $S(\lambda)$) as a function of the wavelength, $\lambda$, of the source via an integral transformation of the form

\begin{equation}
h_s(u,\omega_1,\ldots,\omega_\Omega) = \int\limits_0^{\infty}\, I(u,\lambda,\omega_1,\ldots,\omega_\Omega) \cdot S(\lambda)\, {\rm d}\lambda \; .
\end{equation}

The integral kernel $I(u,\lambda,\omega_1,\ldots,\omega_\Omega)$ describes the instrument and contains the influence of the $\Omega$ instrumental parameters mentioned above. $I(u,\lambda,\omega_1,\ldots,\omega_\Omega)$ may be separated into the contributions from the 
response function, $R$, and the
LSF , $\Lambda$,
which also includes the dispersion relation (linking $u$ and $\lambda$), as

\begin{equation}
I(u,\lambda,\omega_1,\ldots,\omega_\Omega) = \Lambda\left(u(\lambda,\omega_1,\ldots,\omega_\Omega),\lambda\right) \, R(\lambda,\omega_1,\ldots,\omega_\Omega).
\end{equation}

The goal of the internal calibration is to
characterise the differences among all the observations, removing the dependence on the $\Omega$ parameters $\omega_1,\ldots,\omega_\Omega$, 
such that all epoch spectra are described by an integral transformation with 
a kernel 
$I_\mu(u,\lambda)$, not depending on the parameters $\omega_1 ,\ldots, \omega_\Omega$ and identical for all observations.
If this is achieved, 
we considered all observations to be reduced to a common instrument ($\mu$), which we refer to as the mean instrument. 
This term does not however imply a mean in any mathematical sense, and 
it
is not uniquely defined. 
Any kernel independent of the instrumental parameters $\omega_1,\ldots,\omega_\Omega$ can in principle serve as a mean instrument.\par
To remove the dependence on the $\omega_1,\ldots,\omega_\Omega$, we may first separate the continuous parameters into intervals with similar conditions. The boundaries of the intervals can be chosen at (or close to) discontinuities. 
For the time parameter, such boundaries may be the times of decontamination or re-focussing of the telescopes (see Fig. 5 in \citealt{Riello2021}). 
For the AC position parameter, it may be the boundaries between CCDs or stitch blocks of a CCD. 
Each combination of a particular interval of the continuous parameters and the discrete parameters we refer to 
as a calibration unit. Within each calibration unit, the instrument only shows, by construction, a smooth variation.\par
For the moment, we assume that the smooth variations inside a given calibration unit are negligible, and we come back to them later. We may index the calibration units with an index $\kappa$, with some arbitrary but fixed ordering. 
To any index $\kappa$, 
there are intervals in the continuous parameters and particular values of the discrete parameters assigned to it. 
Then, we may link any epoch spectrum in $\kappa^{\rm th}$ calibration unit with the SPD via
\begin{equation}
h_{s \kappa}(u) =  \int\limits_0^{\infty}\, I_{\kappa}(u,\lambda) \cdot S(\lambda)\, {\rm d}\lambda,
\end{equation}
with $I_{\kappa}(u,\lambda)$ being the instrument kernel for the $\kappa^{\rm th}$ calibration unit, where, as previously stated, we neglected the inter-calibration unit variations for the time being and did not consider dependency of the $I_\kappa(u,\lambda)$ on the $\omega_1,\ldots,\omega_\Omega$ parameters.

To be able to internally calibrate the instrument at all, we have to assume that for each calibration unit, 
a
linear transformation exists between the epoch spectra in the $\kappa^{\rm th}$ calibration unit, and the mean internal instrument, $\mu$, which can, but does not necessarily have to be, one of the calibration units. 

Since all linear transformations between well-behaved mathematical functions can be 
uniquely expressed by an integral transform, we may choose to express the transformation between 
the mean instrument $\mu$ and the $\kappa^{\rm th}$ calibration unit as an integral transform, using a transformation kernel $A_{\kappa}$, which thus has the following form:
\begin{equation}
h_{s \kappa}(u) =  \int\limits_{-\infty}^{\infty}\, A_{\kappa}(u,u^\prime) \cdot h_{s \mu}(u^\prime)\, {\rm d}u^\prime . \label{eq:basicTransform}
\end{equation}
For the mean spectrum $h_{s \mu}(u)$, we chose a continuous representation by a linear combination of basis functions, $\varphi_n(u)$ (see Sect.~\ref{sec:source}). 
Thus, for a particular source, we assumed
\begin{equation}
\label{eq:sourcebasisfunctions}
h_{s \mu}(u) = \sum\limits_{n=0}^{N}\, b_{s n} \cdot \varphi_n(u) ,
\end{equation}
with $b_{s n}$ the coefficients of the spectrum of a source $s$ in the finite development with $N+1$ basis functions. We highlight that there is a single set of $b_{s n}$ coefficients for a given source, independently of the epoch and calibration unit of their observations.
Inserting Eq.~(\ref{eq:sourcebasisfunctions}) into Eq.~(\ref{eq:basicTransform}) gives
\begin{eqnarray}
h_{s \kappa}(u) & = &  \int\limits_{-\infty}^{\infty}\, A_{\kappa}(u,u^\prime) \, \sum\limits_{n=0}^{N}\, b_{s n} \cdot \varphi_n(u^\prime) \,{\rm d}u^\prime \\ 
 & = & \sum\limits_{n=0}^{N}\, b_{s n}\, \int\limits_{-\infty}^{\infty}\, A_{\kappa}(u,u^\prime) \cdot \varphi_n(u^\prime) \, {\rm d}u^\prime . 
\end{eqnarray}
We can thus express the epoch spectrum in $\kappa^{\rm th}$ calibration unit as a linear combination of basis functions with the same coefficients used for the mean spectrum, 
and with the basis functions in $\kappa^{\rm th}$ calibration unit being the basis functions chosen for the mean instrument transformed by the application of $A_{\kappa}$.\par
For practical computations, we can approximate the integral in the equation above using Riemann sums, meaning we use the following approximation:
\begin{equation}
h_{s \kappa}(u_i) \approx \sum\limits_{n=0}^{N}\, b_{s n}\, \sum\limits_{j = -\infty}^{\infty}\, A_{\kappa}(u_i,u_j) \cdot \varphi_n(u_j) .
\end{equation}
We may introduce a second approximation, namely that the kernel $A_{\kappa}$ describing the transformation between the mean instrument 
and the $\kappa^{\rm th}$ calibration unit is localised. The scale over which variations in a spectrum are correlated at different $u$ 
is determined by the width of the LSF, and the LSF itself is a localised function. We may therefore achieve a good approximation to
truncate the infinite summation to finite limits around each discrete point $i$. We obtain
\begin{equation}
h_{s \kappa}(u_i) \approx \sum\limits_{n=0}^{N}\, b_{s n}\, \sum\limits_{j = -J}^{J}\, A_{\kappa}(u_i,u_{i+j}) \cdot \varphi_{n}(u_{i+j}) . \label{eq:basicConvolution}
\end{equation}
This expression is essentially a discrete convolution of the mean spectrum to obtain the different epoch spectra (which correspond to different $\kappa^{\rm th}$ calibration units), with 
the generalisation that the discrete convolution kernel $A_{\kappa}(u_i,u_{i+j})$ is not only a function of $u_i - u_{i+j}$, 
but of both $u_i$ and $u_{i+j}$ individually. $A_{\kappa}(u_i,u_{i+j})$ may therefore vary with $u$ during the discrete 
convolution process. We chose Eq.~(\ref{eq:basicConvolution}) as the basic formulation of the internal instrumental calibration 
model for {\Gaia}'s spectra, which consists of determining $A_{\kappa}$ once the set of source basis functions is fixed.\par

\begin{figure}
   \centering
\includegraphics[width=0.49\textwidth]{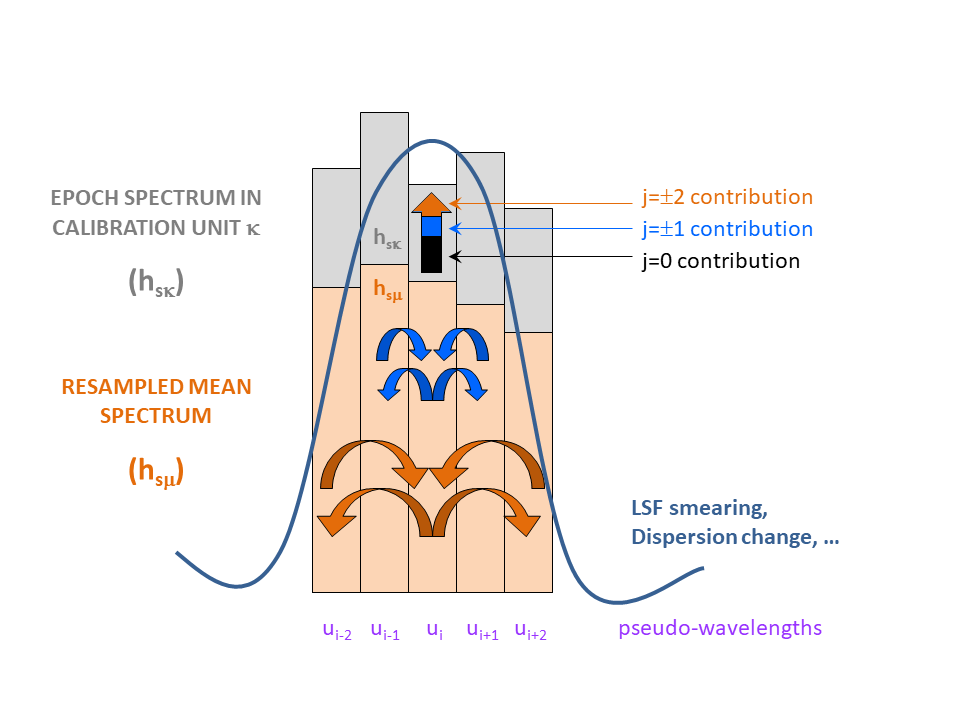}
\caption{
Graphics showing contribution from neighbouring pixels for $J=2$ case in 
Eq.~(\ref{eq:basicConvolution}). This influence from neighbouring pixels needs to be taken into account when predicting the observation $h_{s \kappa}$ in the $\kappa^{\rm th}$ calibration unit starting from the reference mean spectrum $h_{s \mu}$.
  }
  \label{fig:neighb}
\end{figure}

The nature of the discrete convolution with a variable kernel is illustrated in Fig.~\ref{fig:neighb}. For any given sample $u_i$ of the spectrum, the calibrated spectrum results from an exchange of flux between neighbouring samples $u_{i+j}$, where the index $j$ runs over the finite range $-J$ to $J$. The discrete kernel $A_{\kappa}(u_i,u_{i+j})$ describes the amount of exchange of flux between the samples $u_i$ and $u_{i+j}$ when transforming the mean spectra into the epoch spectra.\par

Due to the limitation imposed on the $j^{\rm th}$ interval, the maximum flux influence of the neighbours is limited to $\pm J$. The chosen value for $J$ should be selected based on the expected influence of the neighbouring fluxes. As we see in Sect.~\ref{sec:effects}, the width of the LSF and dispersion difference between calibration units recommend the choice of $J$ to be at least two neighbours or more ($J\geq 2$). 

\begin{figure}
   \centering
\includegraphics[width=0.49\textwidth]{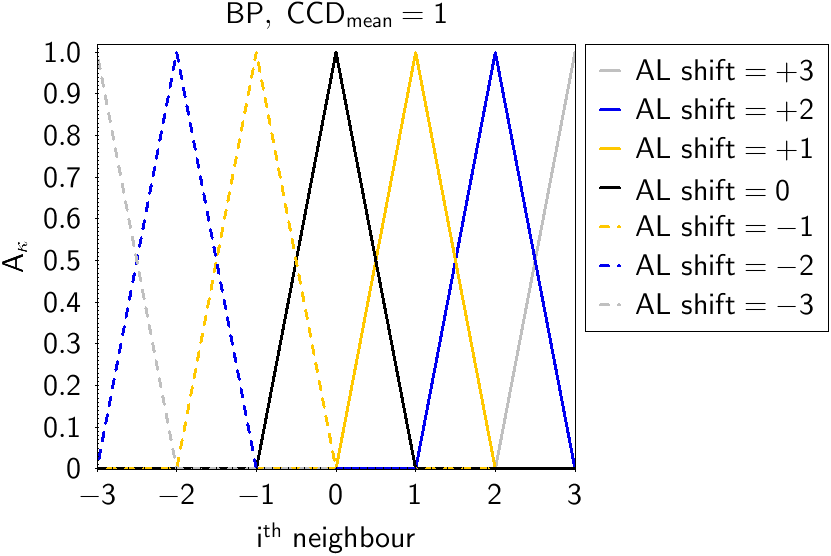}
\caption{Change in the value of the coefficients when different AL shifts (from -3 to +3) are
considered for the observations with respect to the mean spectra. No dispersion and LSF effect variations are considered between the mean spectra and the observation.
  }
  \label{fig:aijshift}
\end{figure}

\begin{figure}
   \centering
\includegraphics[width=0.49\textwidth]{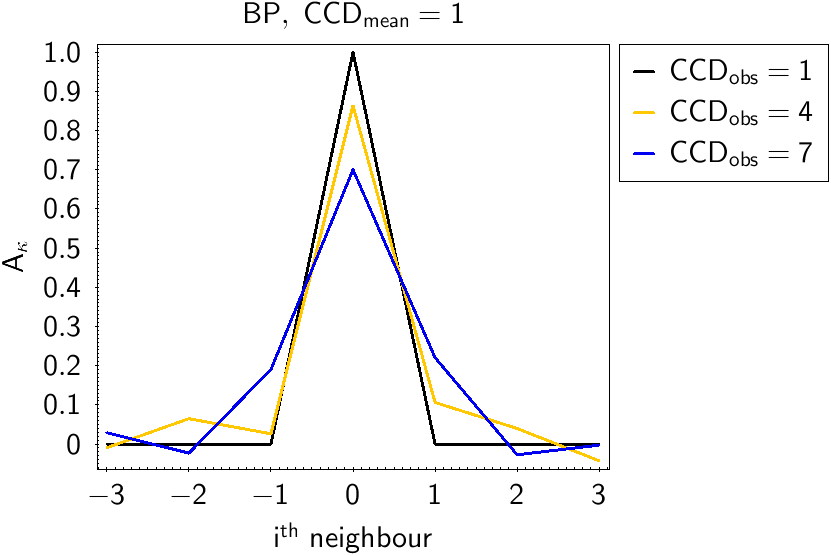}
\caption{Change in the value of the coefficients when different {\Gaia} BP CCDs (with different dispersion and LSF conditions) are
considered for the observations with respect to the mean spectra. The cases of the mean spectra in the CCD1 reference system and observations in CCDs 1 (black), 4 (orange), and 7 (blue) are plotted here.
  }
  \label{fig:aijdisp}
\end{figure}

The choice of $J$ will also limit the maximum AL shift allowed to compare different observations. If some shift is present, the derived $A_{\kappa}$ values will incorporate this effect to account for this shift. For instance, Fig.~\ref{fig:aijshift} shows how the derived values of these coefficients change when the only difference between the mean spectra and the observations are due to AL shifts (no dispersion and LSF changes are introduced in this test and thus only $j=0$ contribute here, $A_{\kappa}=1$, and the rest of neighbours are zero). Although keeping the shape of the $A_{\kappa}$ values was not imposed, the algorithm detects that only a shift of the values is needed to reproduce the shift introduced to the spectra in the AL direction. The range of AL shifts that the model is able to cope with are limited to the $J$ value (the maximum number of neighbours, $J=3$ in this case in Fig.~\ref{fig:aijshift}). 
As one of the previous steps in the pre-processing is the calibration of the CCD geometry (see Fig.~\ref{fig:scheme}), we do not expect to see significant shifts in the internal calibration.
This geometric calibration of the spectra was already done in previous data releases when using the spectrophotometric data to derive the photometric calibration \citep{Carrasco2016}. If instead of producing a shift we perform the prediction of an observation in a different calibration unit in {\Gaia} with different dispersion and LSF conditions among the mean spectra and the observation, this produces a deformation of the $A_{\kappa}$ coefficients (see Fig.~\ref{fig:aijdisp}) with respect to the simple shape seen in Fig.~\ref{fig:aijshift}.

$A_{\kappa}(u_i,u_{i+j})$ is assumed to be a smooth function of $u_i$. For a compact representation of the kernel $A_{\kappa}(u_i,u_{i+j})$, we may choose to represent its dependence on $u_i$ as 
a linear combination of basis functions ($E$). We thus use a development of the following form:
\begin{equation}
A_{\kappa}(u_i,u_{i+j}) = \sum\limits_{l=0}^L\, c_{jl} \cdot E_l(u_i) .
\end{equation}

Since we do not expect abrupt changes of the kernel with $u_i$, a low-order polynomial may be a suitable choice as a basis. In this case, we obtain a parametrisation:
\begin{equation}
A_{\kappa}(u_i,u_{i+j}) = \sum\limits_{l=0}^L\, c_{jl} \cdot \left( u_i - u_{\rm ref} \right)^l  . \label{eq:kernelParameterisation}
\end{equation} 
The value of $u_{\rm ref}$ is a reference pseudo-wavelength that may be chosen in any convenient way (for example, the centre of the window or one of the extremes, once the different spectra have been aligned with the previous geometric calibration procedure). 
This approach reduces the number of coefficients to be fitted and improves the physical interpretation 
of $A_{\kappa}$, reducing degeneracies. \par

Until this point, we neglected the possible smooth variation of the instrument in time and AC position within a particular $k^{\rm th}$ calibration unit. At this point, we are able to take such variations into account. This we do by introducing a polynomial dependence of the $c_{jl}$ coefficients
on the AC position, $x_i$, for the pseudo-wavelength $u_i$ with respect to a reference AC position, $x_{\rm ref}$, as follows: 
\begin{equation}
\label{eq:cjlACposition}
c_{jl}=
\sum\limits_{k=0}^K\, 
d_{jkl} 
\cdot \left( x_i - x_{\rm ref} \right)^k .
\end{equation}
This results in the expression\\
\begin{equation}
A_{\kappa}(u_i,u_{i+j},x_i) = 
\sum\limits_{l=0}^L\, 
\sum\limits_{k=0}^K\, 
d_{jkl} 
\cdot \left( x_i - x_{\rm ref} \right)^k
\cdot \left( u_i - u_{\rm ref} \right)^l . \label{eq:kernelParameterisationAC}
\end{equation}

Combining this equation with Eq.~(\ref{eq:basicConvolution}), we obtain the following expression to derive the prediction of the epoch spectra $h_{s \kappa}$ for $\kappa^{\rm th}$ calibration unit in a pseudo-wavelength $u_i$ and located at the AC position $x_i$,

\begin{eqnarray}
 \label{eq:basicConvolutionFinal}
h_{s \kappa}(u_i,x_i) &\approx& \sum\limits_{n=0}^{N}\,  \sum\limits_{j = -J}^{J}\, \sum\limits_{k=0}^K\, \sum\limits_{l=0}^L\, 
b_{s n}
\cdot d_{jkl} 
\cdot \left( x_i - x_{\rm ref} \right)^k \cdot \\
&&
\cdot \left( u_i - u_{\rm ref} \right)^l
\cdot \varphi_{n}(u_{i+j}) \nonumber . 
\end{eqnarray}

Time variations of the instrument are dealt with by solving the instrument model in specific time intervals, meaning we used observations taken within a given time range. These time ranges are treated as another calibration unit with different instrumental conditions. The lengths of these intervals need to be longer when calibrating gated and 2D observations due to the limited number of observations obtained in these configurations. 

We refer to the process of determining the instrumental coefficients, $d_{jkl}$, as an 'instrument update', keeping the coefficients of the calibration sources in the representation of the mean spectra, $b_{s n}$, fixed. On the other side, the mean spectra of the source are known when $b_{s n}$ coefficients are determined. We refer to the derivation of these $b_{s n}$ coefficients for all calibration sources as 'source updates', keeping the instrumental parameters, $d_{jkl}$, fixed (see Fig.~\ref{fig:scheme}). Once the instrumental coefficients are known with the calibration sources, the same $d_{jkl}$ values can be applied to the rest of sources (non-calibrators).

\section{Instrument calibration}
\label{sec:instrumentcalibration}

The determination of the instrument coefficients, $d_{jkl}$ in Eq.~(\ref{eq:basicConvolutionFinal}),
requires the knowledge of the $b_{s n}$ coefficients 
representing the sources as in the mean instrument. Ideally, these coefficients are known for $M$ calibration sources, with $M$ being sufficiently large (of the order of millions for the {\Gaia} case). We refer to these sources as the 'internal calibrators'.
However, before the instrument calibration is performed, the representation of sources in the mean instrument is unknown. To overcome this deadlock, an iterative approach was adopted. We may assume either the coefficients of the internal calibrators in the mean instrument ($b_{s n}$) to be known and derive the parameters describing the transformations to the different calibration units ($d_{jkl}$), or we may assume the opposite scenario, in which the instrumental kernel to transform between the different calibration units is known, and derive the source coefficients of the internal calibrators in the mean instrument. These steps can be applied iteratively, and the convergence can be monitored from the agreement between the epoch spectra of the internal calibrators and their predictions, derived from the mean spectra and using the transformation obtained between the mean instrument and the epoch spectra.\par
Starting this iterative process requires some initial values, either for the parameters describing the transformations, or the mean spectra of the internal calibrators. The simplest approach is to assume no differences between the different calibration units, hence using the identity transform between the mean instrument and all calibration units: 
\begin{equation}
d_{jkl}=\delta_{j0}\, \delta_{k0}\, \delta_{l0}
\end{equation}
\noindent with $\delta_{\alpha \beta}$ being the Kronecker delta. This is the behaviour shown in Fig.~\ref{fig:aijshift} when no change (in shift, dispersion, LSF, etc.) is considered between the mean spectra and the observation scales. 
With this initial guess for the instrument parameters, Eq.~(\ref{eq:basicConvolutionFinal}) can be solved for the coefficients $b_{s n}$ for all internal calibrators. In a following step, the $b_{s n}$ can be kept fixed, and Eq.~(\ref{eq:basicConvolutionFinal}) can be solved to obtain the instrumental parameters $d_{jkl}$. 
The iterative process of source update and instrument update can continue until convergence is reached. 

To ensure convergence, two additional considerations can be introduced.
First, the epoch spectra predicted by the model remain the same if the source coefficients, $b_{s n}$, are multiplied by an arbitrary scaling factor, and the 
instrumental coefficients, $d_{jkl}$, are
divided by the same scaling factor. This effect corresponds to a degeneracy between the instrumental response and the flux scale of the mean spectra. If this effect is producing a divergence of the solution, this can be mitigated through a re-normalisation of the instrumental parameters after every instrument update. This can be introduced by dividing the instrumental kernel by their sum over all neighbours, AC positions, and calibration units. 

Secondly, all $\kappa$ calibration units are required to converge to the same mean instrument; that is, systematic differences between the mean spectra of two sources with identical SPDs but a different distribution of their observations in calibration units need to be avoided. This can be achieved by employing sources with observations in several different calibration units. One example of this is the use of sources observed with different gating conditions. This kind of source will ensure the convergence of the solution for the different gates to a common reference system without sudden jumps in the behaviour of the calibrated instrument with magnitude. Further considerations on the selection of the internal calibrators are included in Sect.~\ref{sec:calibrators}.


An additional complication of the internal calibration process results from the observation of a window of finite size around the expected position of a source on the CCD. This limited size of the window added to the inaccurate centring of the source within the window and AC motion (as discussed in Sect.~\ref{sec:effects}), means that the observed spectrum only includes a fraction of the flux of a particular source.
A correction for the flux loss is included within the internal calibration procedure.\par
The source can be de-centred within the window by a certain amount in AL and AC directions ($\Delta y$ and $\Delta x$, respectively). We evaluated, using pre-flight simulations, the expected contribution of the flux loss due to AL and AC shifts, the colour of the source ({\BP}$-${\RP}), and AC motion ($v_{AC}$). 
Based on these results, we can apply a model splitting the contributions 
to the flux loss into different components:

\begin{equation}
\label{eq:fluxloss}
\frac{h_{s \kappa}(u_i,x_i,\Delta x, \Delta y, G_{\rm BP}-G_{\rm RP}, v_{AC})}{h_{s \kappa}(u_i,x_i,0, 0, G_{\rm BP}-G_{\rm RP}, 0)} =  
1 + P
+ Q ,
\end{equation}

\noindent with $h_{s \kappa}(u_i,x_i,\Delta x, \Delta y, G_{\rm BP}-G_{\rm RP}, v_{AC})$ being the flux once the losses have been accounted for 
and $P$ and $Q$ the AL and AC flux loss contribution, respectively: both of them defined as negative 
and represented here by two polynomial functions (Eqs.~\ref{eq:ALfluxloss} and \ref{eq:ACfluxloss}). 
$h_{s \kappa}(u_i,x_i,0, 0, G_{\rm BP}-G_{\rm RP}, 0)$ is the calibrated flux obtained from Eq.~(\ref{eq:basicConvolutionFinal}) without considering flux loss.

We used a polynomial representation for both $P$ and $Q$ functions on their dependences.
The AL flux loss, $P$ term in Eq.~(\ref{eq:fluxloss}), does not depend on the AC motion ($v_{AC}$):

\begin{equation}
\label{eq:ALfluxloss}
P(G_{\rm BP}-G_{\rm RP}, \Delta y) = \sum_{\alpha=0}^{N_{\alpha}} \sum_{\beta=0}^{N_{\beta}} p_{\alpha\beta} \cdot 
(G_{\rm BP}-G_{\rm RP})^\alpha \cdot (\Delta y)^\beta ,
\end{equation}

\noindent but the AC flux loss does:

\begin{equation}
\label{eq:ACfluxloss}
Q(G_{\rm BP}-G_{\rm RP},\Delta x,  
v_{AC}) =
\sum_{\alpha=0}^{N'_{\alpha}} 
\sum_{\gamma=0}^{N_\gamma} 
 \sum_{\nu=0}^{N_{\nu}} 
q_{\alpha\gamma\nu}
(G_{\rm BP}-G_{\rm RP})^\alpha
(\Delta x)^\gamma 
(v_{AC})^\nu .
\end{equation}

$N_{\beta}$, $N_{\gamma}$ and $N_{\nu}$ are, respectively, the degree chosen for the dependence on AL and AC centring error and AC motion.
$N_{\alpha}$ and $N'_{\alpha}$ represent the degree for the polynomial dependence on the colour for the AL and AC flux loss terms, respectively.
The typical values for the maximum degree of dependence with all these parameters are not expected to be larger than first or second order polynomials, according to the evaluations of these effects explained in Sect.~\ref{sec:effects}.
Although we mention in Sect.~\ref{sec:effects} that the AC flux loss dependence with colour is not very critical (almost grey compared with 
the dependence on the AL flux loss) and that most of the colour dependence is accounted for in the AL flux loss equation, Eq.~(\ref{eq:ALfluxloss}), we prefer to keep the colour dependence in the AC flux loss model, Eq.~(\ref{eq:ACfluxloss}), to account for its small dependence if needed, but it could be neglected (setting $N'_\alpha=0$) if this dependence is found to not be important.

\section{Representation of the mean spectra}
\label{sec:source}

To choose the basis functions $\varphi_n(u)$, $n=0,\ldots,N$ to describe the mean spectra in Eq.~(\ref{eq:sourcebasisfunctions}), two approaches are possible in principle. One is the construction of empirical basis functions, the other one is the use of a generic set of basis functions. The second option is more suitable for the calibration approach described here. First, because the mean instrument is not defined in the initial state of instrument calibration. And second, because the basis functions chosen to describe the spectra in the mean instrument should not be limited by the choice of a subset of spectra used for the construction of an empirical set of bases. 
Once the set of generic basis functions is defined and used, they can always be improved to minimise the number of needed functions and increase their suitability to describe any type of spectra. 

As generic basis functions, a classical orthonormal set of bases provide suitable choices. For a higher level of efficiency, and thus to be able to represent any spectrum with a small number of basis functions, $N$, we may choose a set of bases that already resemble the basic features of the spectra to be represented. This makes polynomial bases less attractive. 

One type of basis functions explored in tests with simulations in Sect.~\ref{sec:testsimu} were the S-splines, also used in {\Gaia} for the PSF fitting as explained in \cite{Rowell2021}. Other types of functions that can be used are the Hermite functions (see tests with {\Gaia} observations in Sect.~\ref{sec:testdata}). When considering Hermite functions, the two first ones are as follows:
\begin{eqnarray}
\varphi_0(\theta) & = & \pi^{-\frac{1}{4}}\; {\rm e}^{-\frac{\theta^2}{2}} \quad ,\\
\varphi_1(\theta) & = & \sqrt{2}\, \pi^{-\frac{1}{4}}\; \theta \, {\rm e}^{-\frac{\theta^2}{2}} \quad .
\end{eqnarray}
The other following Hermite functions satisfy the recurrence relation:
\begin{equation}
\varphi_{n}(\theta) = \sqrt{\frac{2}{n}}\, \theta \, \varphi_{n-1}(\theta) - \sqrt{\frac{n-1}{n}} \, \varphi_{n-2}(\theta) \quad ,
\end{equation}
allowing us to compute any higher order Hermite function. Basically, the Hermite functions arise from Hermite polynomials by multiplication with a Gaussian function. As a result, they are centred at $\theta=0$, and for sufficiently high $\theta$ absolute value, all Hermite functions are converging to zero. This resembles the global behaviour of {\Gaia}'s spectrophotometry, which is also converging to zero for focal plane positions sufficiently far from the location of the source.\par
Hermite functions are furthermore orthonormal with respect to each other, meaning they satisfy the following orthonormality condition:
\begin{equation}
 \int\limits_{-\infty}^{\infty}\, \varphi_n(\theta) \cdot \varphi_m(\theta) \, {\rm d}\theta = \delta_{nm} \; .
\end{equation}
As the Hermite functions are centred on $\theta=0$, the origin of the pseudo-wavelength axis, $u$, should be shifted such that the origin becomes close to the centre of the spectra. Furthermore, the pseudo-wavelength axis needs to be re-scaled such that the width of the spectra is met to the width of Hermite functions. We therefore consider a linear transformation between the pseudo-wavelength axis $u$ and the argument of the Hermite functions $\theta$ of the form
\begin{equation}
\theta = \Theta\cdot u + \Delta\theta \; .
\end{equation}
The scaling parameter $\Theta$, the shift $\Delta\theta$, and the number of basis functions $N$ need to be determined based on actual spectra. The three parameters may, however, not be chosen independently of each other. If a spectrum has to be represented over a certain range of pseudo-wavelengths, then choosing a smaller scaling parameter $\Theta$ results in the Hermite functions expanding outwards less in pseudo-wavelengths. Accordingly, to cover the same pseudo-wavelength interval, the number of Hermite functions needs to be increased. Thus, smaller $\Theta$ imply larger $N$, and vice versa. In our case, to optimise the choice of $\Theta$ and $N$, combinations of both such that the outermost local extreme of the highest order Hermite function is closest to the boundary of the pseudo-wavelength interval of interest have been investigated. In the {\Gaia} case, a maximum of 30 samples at each side to account for the total size of the window should be chosen. The combination of $\Theta$ and $N$ with the lowest level of residuals has to be chosen and adopted.\par
\begin{figure}[!htbp]
   \centering
\includegraphics[width=0.9\columnwidth]{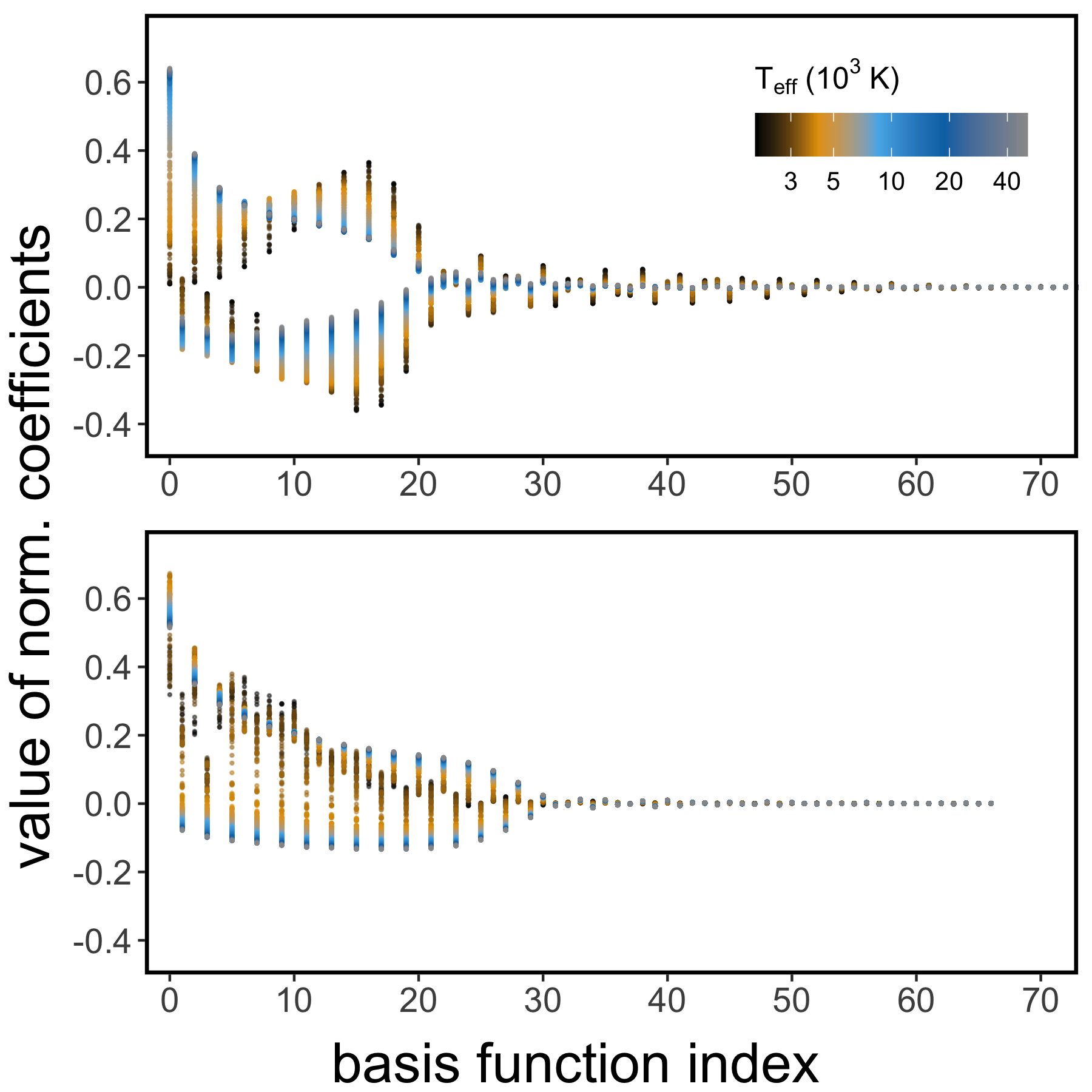}
\caption{Normalised coefficients of simulated spectra for BP (top panel) and RP (bottom panel), for all BaSeL spectra \citep{Lejeune1997} with solar metallicity. The colour scale indicates the effective temperature ($T_{\rm eff}$).}
  \label{fig:bnteff}
\end{figure}

An example for the representation of {\Gaia} spectra with Hermite functions is shown in Fig.~\ref{fig:bnteff}. The BP and RP spectra of all model SEDs of the BaSeL spectral library \citep{Lejeune1997} with solar metallicity were simulated and developed in Hermite functions. The coefficients for BP and RP are shown, normalised with respect to the $l_2$-norm, and the effective temperature ($T_{\rm eff}$) is indicated. The two branches of the coefficients visible in this figure correspond to even (larger coefficients) and odd (smaller coefficients) Hermite functions. Stars with different effective temperatures show systematic differences in the distribution of the coefficients with the order of the Hermite functions. A discrimination of different stellar spectra is therefore possible using the representation in Hermite functions, without explicitly sampling the spectra as a function of pseudo-wavelength, avoiding the loss of information in this process, as the set of coefficients have the most complete information available.

\begin{figure*}[!htbp]
   \centering
\includegraphics[width=0.49\textwidth]{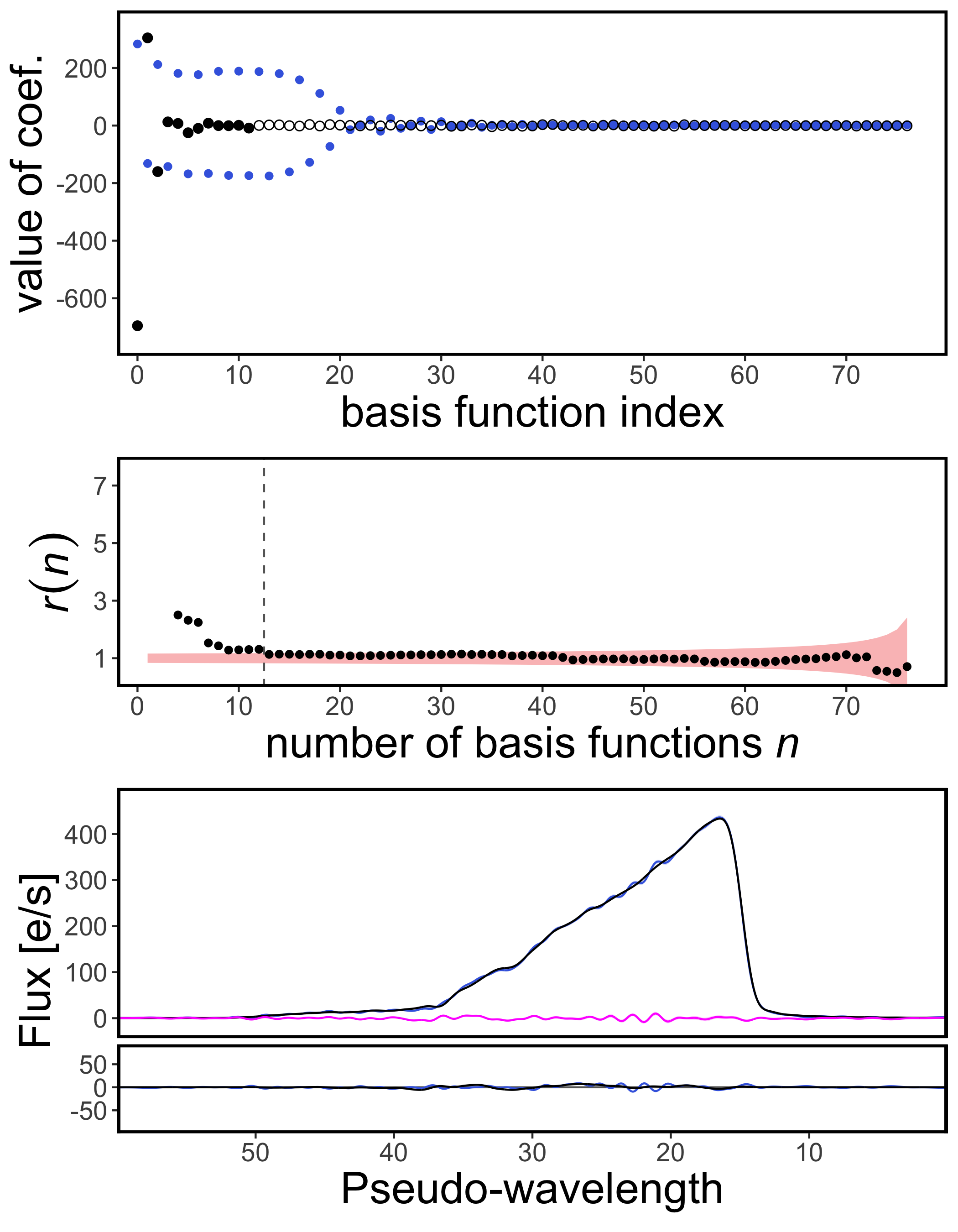}
\includegraphics[width=0.49\textwidth]{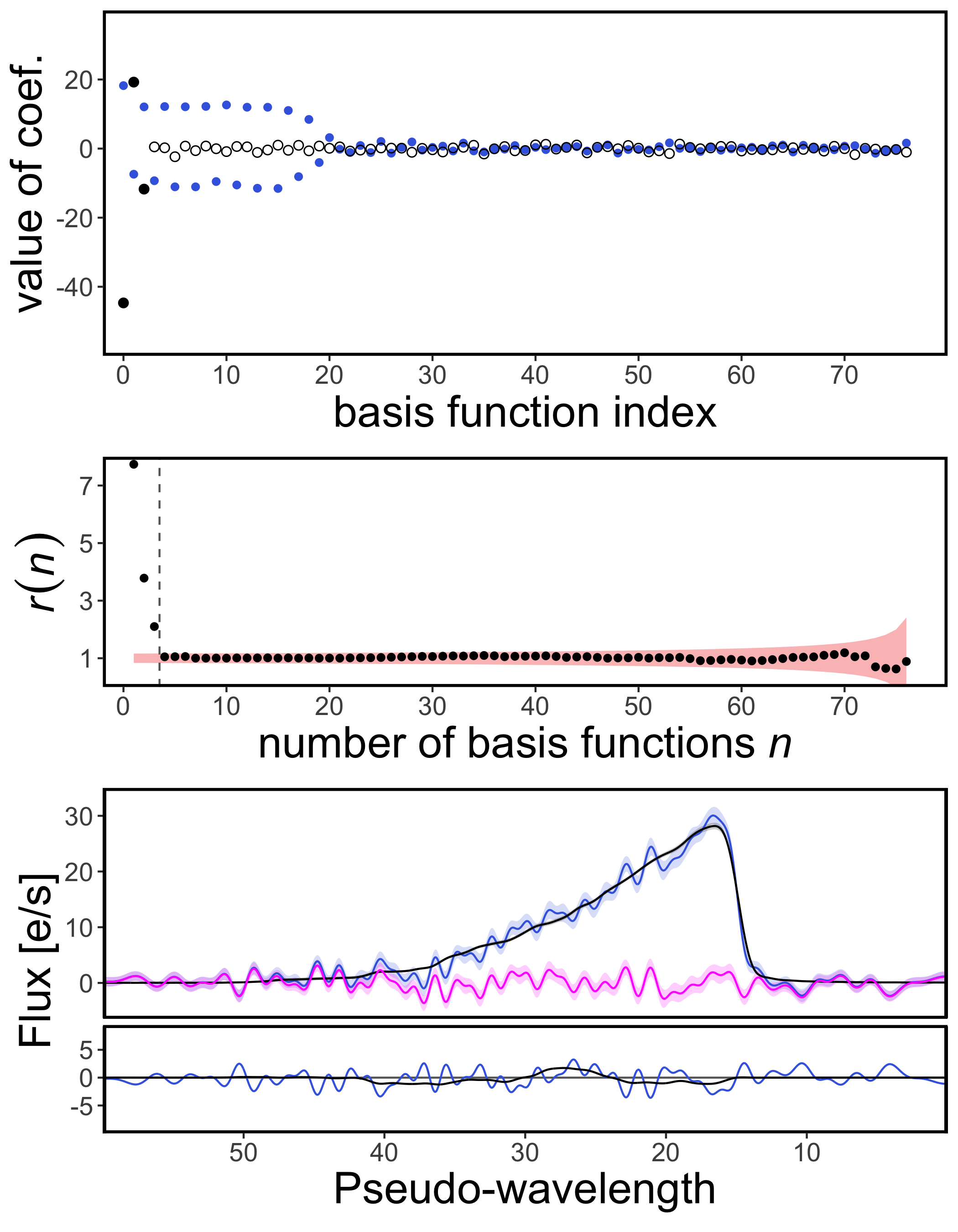}
\caption{Top panel: Coefficients that describe a BP spectrum for a simulated Sun-like star, with $G=16$~mag (left) and $G=19$~mag (right). Blue symbols are the coefficients in Hermite basis functions, black symbols represents those in rotated basis functions. Open symbols represent the coefficients dropped after truncation. Middle panel: Normalised standard deviation of the coefficients, $r(n)$, when excluding basis functions in ascending order. The red shaded region indicates twice the standard error on the standard deviation. $r(n)$ values smaller than this threshold could be ignored producing similar mean spectra. Bottom panel: Spectra as a function of pseudo-wavelength, for all coefficients (blue line), and for the truncated representation in the rotated Hermite basis functions (black line). The magenta line shows the contribution lost due to truncation. At the bottom of the panel, the differences with respect to the correct representation of the spectra are shown. Shaded regions indicate the 1-$\sigma$ confidence regions.
An approximate correspondence between pseudo-wavelength values and absolute wavelengths can be found in Fig.~\ref{fig:wavelength}.
}
  \label{fig:shrinking}
\end{figure*}

As a generic set of bases for functions on the real axis, S-splines or Hermite functions allow for a good 
representation of spectra. They are, however, not efficient in doing so, as they have no deeper relation to {\Gaia} spectra. Consequently, a large number of basis functions may be required to represent spectra. 
As a reference, we now consider the use of Hermite functions.
A more efficient set of bases (in terms of number of functions) can, however, be constructed 
from Hermite functions 
by finding suitable linear combinations thereof that more efficiently represent the features present in the spectra. 
To find such linear combinations of Hermite functions,
 we may select a set of spectra for 
BP and RP, respectively, whose spectra we consider 'typical' for the majority of sources 
in the sky. In general, these will be sources of intermediate colours. We may select sources 
with good signal-to-noise ratios, that is, medium to bright sources, and normalise their coefficient 
vector representing them in Hermite functions with respect to their $l_2$-norm. We arrange the $N$ 
normalised coefficients for $N_{\rm src}$ sources in an $N_{\rm src} \times N$ matrix $\bf C$, 
which we subject to a singular value decomposition (SVD). The orthogonal matrix $\bf V_C$ 
from this SVD represents a rotation of the Hermite functions to a new set of basis functions (Fig.~\ref{fig:shrinking}) concentrating the relevant information into the first basis functions in order to represent typical spectra. This approach corresponds 
to the one used by \cite{Weiler2020} for adjusting basis functions to a set of calibration 
sources. The only difference is that we applied this approach here to obtain a representation of 
{\Gaia} BP and RP spectra.

As the most important information in the rotated optimised basis functions is included in the first coefficients, this new representation can be exploited to allow truncating the number of source coefficients. In order to reduce the number of basis functions, we could keep only those coefficients significantly larger than their uncertainties. An example of this optimisation and truncation process is shown in Fig.~\ref{fig:shrinking}. The derived coefficients for simulated BP spectra of a Sun-like star using Hermite functions are represented as blue symbols in the top panels of Fig.~\ref{fig:shrinking}. The SPD was taken from the BaSeL spectral library \citep{Lejeune1997}, with the parameters $T_{\rm eff}=5750$~K, $\log g=4.5$ and [M/H]$=0.0$, and the simulations were done assuming a focused {\Gaia} telescope, as described by \cite{Weiler2020}. The left-hand side shows the results for $G=16$~mag, the right-hand side for $G=19$~mag. The two branches of the coefficients correspond to even (positive coefficients) and odd (negative coefficients) Hermite functions. The resulting spectra using these coefficients are shown as blue lines in the bottom panels of Fig.~\ref{fig:shrinking}. Black circles in the top panels show the same representation in a linear combination of Hermite functions, which results from an orthogonal transformation derived to optimise the representation of BP spectra for all BaSeL SPDs. Both sets of coefficients, the original one and the transformed one, describe exactly the same BP spectrum. In the transformed representation, coefficients with values significantly different from zero (beyond their uncertainties) only occur for the leading basis functions (filled black circles in top panels). This opens the possibility of ignoring any basis functions other than the leading ones, thus significantly reducing the amount of data required for representing the BP spectrum and also its noise. The resulting spectra are represented in the bottom panels of Fig.~\ref{fig:shrinking} as a black line. This process only removes the coefficients that are indistinguishable from noise, meaning that no relevant information is lost in the process.

The decision of how many leading coefficients are required for the representation of the spectrum is based on the ratio of the value of the coefficient over the square root of its variance. We considered only the first $n$ coefficients, and computed the standard deviation of the remaining coefficients (from $n+1$ to $N$) over the square root of their variances, which we denote $r(n)$,
\begin{equation}
r(n) \coloneqq {\rm sd}\left(\, \frac{b^{\prime}_{s\ n+1}}{\sigma^{\prime}_{s\ n+1}},\ldots , \frac{b^{\prime}_{sN}}{\sigma^{\prime}_{sN}}\, \right) \quad .
\end{equation}
Here, the $b^{\prime}_{sn}$, with
\begin{equation}
{\bf b}^{\prime}_s = {\bf V_C}\, {\bf b}_s \quad ,
\end{equation}
denote the coefficients in the rotated Hermite basis, and the $\sigma^{\prime}_{sn}$ the corresponding square roots of the variances of the coefficients in this basis. As correlations between coefficients are low in the rotated Hermite basis, we neglected them in the following. The quantity $r(n)$ has a standard error of
\begin{equation}
\sigma_r(n) = \frac{1}{\sqrt{2\, (N - (n+1))}} \quad ,
\end{equation}
which is larger when $n$ increases, as seen in the red shaded area of the central panel in Fig.~\ref{fig:shrinking}, because there are fewer elements remaining.
We chose a threshold $a$, and take the first $n^\prime$ coefficients such that $n^\prime$ is the largest number with 
\begin{equation}
r(n^\prime) > 1 + a \cdot \sigma_r(n^\prime).
\end{equation}
The coefficients corresponding to basis functions with indices larger than $n^\prime$ can be neglected. This procedure is illustrated in the central panel of Fig.~\ref{fig:shrinking}. The black symbols show the $r(n)$ values, while the shaded region shows the range of $1 \pm \, a \cdot \sigma_r(n)$, for the threshold $a = 2$. The cut in the number of leading coefficients is indicated by the dashed line. 

The chosen criterion depends on the magnitude of the source considered. For fainter sources, the noise is larger, and the number of relevant coefficients decreases. In the example shown in Fig.~\ref{fig:shrinking}, we obtain $n^\prime = 12$ for $G = 16$~mag, and $n^\prime = 3$ for $G = 19$~mag, for the otherwise same SPD. The resulting spectra as a function of pseudo-wavelength are shown as black lines in the bottom panel of Fig.~\ref{fig:shrinking}. The magenta lines show the contribution from the rejected coefficients to the representation of the spectrum. In the bottom part of that panel, the differences between the spectra represented by all coefficients, and represented by the first $n^\prime$ coefficients in the rotated Hermite basis and the true, noiseless representation of the spectra, are shown. A good representation of the spectrum can thus be achieved with a considerably smaller number of coefficients. The contribution of the neglected coefficients essentially introduces noise to the representation of the spectra.

This truncation process in the number of coefficients has the potential of significantly reducing the size of the catalogue, which is very convenient in the case of {\Gaia} due to the large amount of sources observed. Thus, any strategy to reduce the number of coefficients to be determined (and consequently also reducing the size of the covariance matrices) but keeping the same physical information can be extremely useful.
The actual compression in the amount of data that can be achieved this way depends strongly on the choice of the reference set of sources that are used for deriving the orthogonal transformation to an optimised basis, and on the noise of the spectra to which the optimisation is applied. As explained, faint sources with large noise can be represented by a smaller number of basis functions, as can sources that have spectra that are similar to the reference set of sources.

The truncation scheme proposed in this section can be used to reduce the impact of noise when analysing the spectrophotometric data. For instance, this is done in some modules of the algorithms used to determine the astrophysical parameters inside {\Gaia} Collaboration (by the astrophysical parameters determination unit, CU8). In other modules where the spectral resolution is more critical (like analysis of emission lines in the study of quasars, for instance), CU8 prefer to use the whole set of coefficients including those with the noise contribution. When using all coefficients beyond the noise level, one has to be aware that the fine details in the spectra that could be confused with real spectral lines, could simply be due to the effect of noise in the coefficients. The exact process developed by CU8 will be described in a forthcoming paper accompanying {\GDR3}.

The optimisation of the source basis functions can be done at the end of the whole calibration and source update processes. Thus, when in the following sections we talk, for instance, about Hermite functions, the reader should take into account that they will not be the final functions, but a combination of them, as described in this section.

\section{Selection of the calibration sources}
\label{sec:calibrators}

The instrumental calibration is done using a set of 'well-behaved' calibration sources (of the order of a few million) 
assumed to be non-variable
during the time span of the observations. 
If the observed spectra for these calibration sources differ with respect to what is predicted from the knowledge of the source, the true instrument for the given epoch of the observation differs from the mean instrument. We define a well-behaved source as a {\Gaia} source with a sufficient number of observations in good conditions (no readout failures, no dead pixels, charge injections, etc.). 
Calibration sources need to be isolated to make sure that their spectrum is not disturbed by neighbouring sources falling within the boundary of the recorded window \citep{Riello2021}.
Moreover, we should also avoid flux contamination
from any nearby bright source, the long tails of which can influence sources at large
separations (see Figs. 3 and 4 in \citealt{Torra2021}). To produce an instrument calibration valid for all types of sources, we should ensure that the set of calibration sources 
contains all ranges of magnitudes and colours.

The basic problem of the choice of the internal calibrators can be made visible by re-formulating Eq.~(\ref{eq:basicConvolution}). Assuming we have the $M$ internal calibrators and suppress the index $\kappa$ for the calibration unit for the moment, for each of the $M$ internal calibrators, indexed $m$, $m=1,\ldots,M$, Eq.~(\ref{eq:basicConvolution}) then becomes
\begin{equation}
h_{m}(u_i) \approx \sum\limits_{n=0}^{N}\, b_{mn}\, \sum\limits_{j = -J}^{J}\, A(u_i,u_{i+j}) \cdot \varphi_{n}(u_{i+j}) .
\end{equation}
The last sum on the right-hand side is independent of $m$, it is just the transformed basis function at $u_i$. We may write $\bar{\varphi}_n(u_i)$ for it, and thus obtain the following:
\begin{equation}
h_{m}(u_i) \approx \sum\limits_{n=0}^{N}\, b_{mn}\, \bar{\varphi}_{n}(u_{i}) .
\end{equation}
We may turn this into a matrix equation if we arrange the left-hand side into an $M \times 1$ matrix $H(u_i)$, and the right-hand side in an $M \times N$ matrix containing the $N$ coefficients for the $M$ internal calibrators, named $B$, and the $N \times 1$ matrix $\Phi(u_i)$ with the transformed basis function values at $u_i$, giving
\begin{equation}
H(u_i) = B \cdot \Phi(u_i) .
\end{equation}
For many sampling points $u_i$, we can expand the sizes of $H$ and $\Phi$ from $M \times 1$ and $N \times 1$ to $M \times U$ and $N \times U$, if we have $U$ sampling points $u_i$. The problem, however, is always the same: given the observed internal calibrator spectra in $H$ and the coefficients of the mean spectra for all internal calibrators in $B$, there is a unique solution for $\Phi$, the transformed basis functions, only if $B$ has full rank. This means there have to be as many linearly independent internal calibrator spectra as there are basis functions, otherwise the transformation between epoch spectra and mean spectra is degenerated. The fact of the spectra not being a linear combination of the spectra of the internal calibrators may have systematic errors in their internal calibration for a given calibration unit. From this, it follows that the internal calibrators should have as many linearly independent spectra as possible. Even if the number of spectra is not $N$, it should contain basically all possible astrophysical types of spectra. In this latter case, the degeneracy has little practical impact.

The set of internal calibrators must be populated as homogeneously as possible to avoid optimising 
the calibration for the most common type of objects. If the flat distribution is not possible, then a weighting system 
should be introduced to compensate the results.
For faint sources of extreme colours, there will be an additional problem, as extremely red sources will
have an extremely low flux at blue wavelengths.
Therefore, no precise colour information will exist for these sources, or it will be only available after a high number of observations are obtained.
Red sources are more abundant at the faint magnitude regime than bluer ones, but the same could happen, in principle, for extremely blue sources with very little flux at red wavelengths.

It would be desirable for the calibrations performed in a given area of the sky to be comparable
with the calibrations done in another sky location. To ensure this, we should try to use the same kind
of calibration sources in different sky coordinates if possible (different calibration units covering periods of time surveying different regions of the sky).
Nevertheless, the distribution of these calibration sources in the sky might be intrinsically far from homogeneous, and some 
relative weighting dependent on colour and magnitude is needed to ensure homogeneous results in all calibration units.

In practice, with so many calibration units to account for in the instrument (with a large focal plane and observations collected with different instrumental configurations),
it can be difficult to have enough observations available to calibrate all of them regularly with a short time basis. In this case, the time interval for these calibration units with shortage of observations has to be extended.
Another complementary approach is to use the large- and small-scale strategies, as was done for {\Gaia} photometry (see \citealt{Carrasco2016,Evans2018} and \citealt{Riello2021}). This approach consists of more frequently calibrating large-scale effects for which more observations are available, while small-scale effects can be calibrated over longer time intervals to collect more calibration data.
If these approaches do
not completely solve the shortage of observations available for some calibration units or types of sources
one should relax
the requirements for the calibration sources and 
eventually accept all kind of sources and let the average behaviour for most of the sources dominate the solution.

\section{Application to {\Gaia}}
\label{sec:results}

Although a full statistical analysis of the results will be published when releasing the {\Gaia} spectrophotometric data in {\GDR3} (De Angeli et al, in preparation), in this section we include some results of the derived mean spectra and instrumental coefficients to help the reader understand the kind of output expected 
from this model. 
The calibration approach outlined in this work was tested on simulated data (see Sect.~\ref{sec:testsimu}), as well as with a large subset of real observations (see Sect.~\ref{sec:testdata}).
The shape of the source coefficients are shown and discussed in previous sections (see Sect.~\ref{sec:basicformulations} and \ref{sec:source}) and are not discussed here again.

\subsection{Tests using simulations}
\label{sec:testsimu}
\begin{figure*}[!htbp]
   \centering
\includegraphics[width=0.49\textwidth]{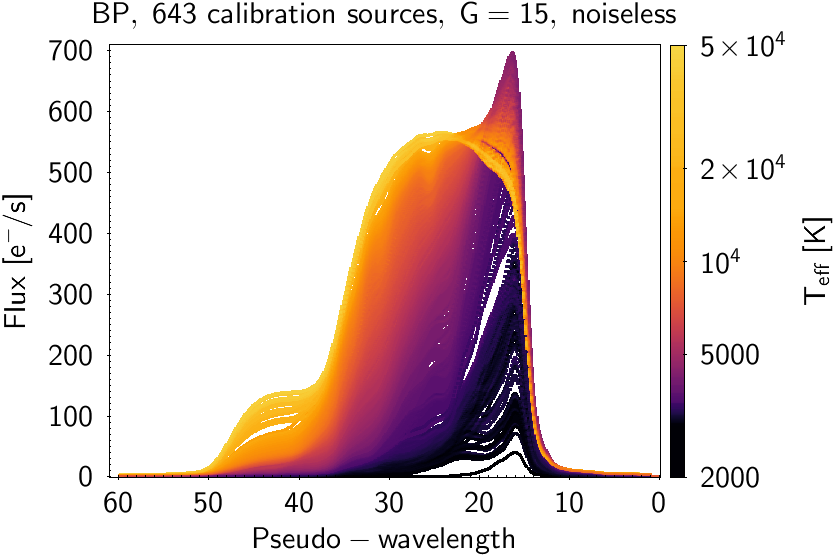}
\includegraphics[width=0.49\textwidth]{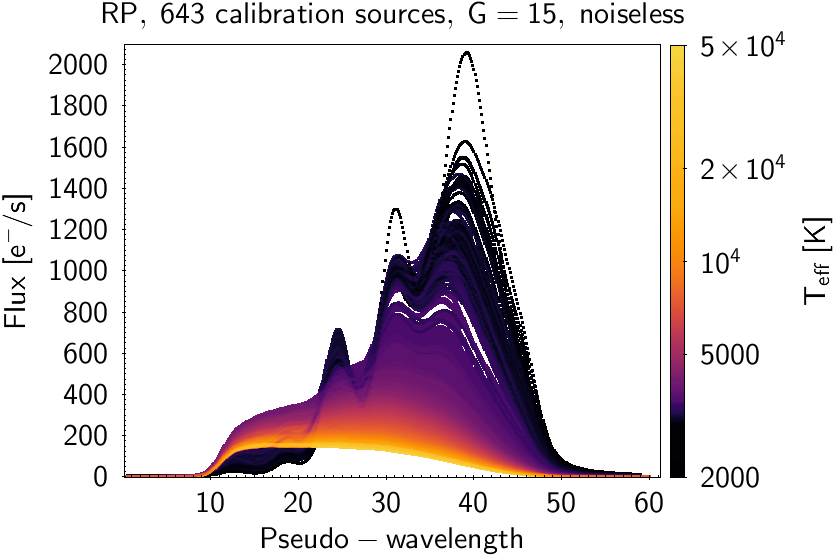}
\caption{CCD7 simulated {\Gaia} spectrophotometry for the 643 sources used as calibrators in BP (left) and RP (right). Colour index indicates the effective temperature of the star. We note that, in RP, the wavelength increases when the pseudo-wavelength does, but the opposite is true in BP. An approximate correspondence between pseudo-wavelength values and absolute wavelengths can be found in Fig.~\ref{fig:wavelength}.
  }
  \label{fig:643sources}
\end{figure*}

For the simulations, we used 643 calibration sources with spectral energy distribution extracted from the library of \cite{Lejeune1997} (see Fig.~\ref{fig:643sources}). These sources cover the ranges $2950<T_{\rm eff}<50\ 000$~K, $0.7<\log g<5.0$~dex, and solar metallicity. 
The BP and RP spectra for these sources were simulated using the GOG simulator \citep{luri2014}, assuming a flux level corresponding to the {\Gaia} magnitude $G=15$~mag. For those tests where we include noise in the observations, we used the uncertainties as derived by GOG simulator (we refer to \citealt{luri2014} for a complete description of the noise model used there).

We distributed eight differently sampled observations in each of the seven BP and RP CCDs. This produced a set of 56 observations per source for each BP and RP with quantum efficiencies, dispersions and LSF conditions.  All observations were simulated using the preceding telescope and not including time variations. Each CCD is treated as a separate calibration unit, as no intra-CCD effects were introduced\footnote{It was tested that the increase of residuals due to intra-CCD sensitivity effects is one order of magnitude smaller than the effects of dispersion and LSF.}. The calibration model used considers the influence of three neighbours, $J=3$ in Eq.~(\ref{eq:basicConvolutionFinal}), and a third-order polynomial representation with the pseudo-wavelength position, $L=3$ in Eq.~(\ref{eq:basicConvolutionFinal}). The reference pseudo-wavelength, Eq.~(\ref{eq:kernelParameterisation}), was fixed to $u_{\rm ref}=30$. As this experiment does not consider intra-CCD variations, no smoothing on AC position was considered, $K=0$ in Eq.~(\ref{eq:basicConvolutionFinal}). This gives a total of 28 instrumental coefficients, $d_{jkl}$, to be fitted. The source basis functions used were S-splines\footnote{The definition of these functions can be found in the {\GEDR3} documentation (Section 3.3.5) and in \cite{Rowell2021}.}, using $N=137$~knots. Only the central 40 positions were fitted, as the wings of the spectra correspond to wavelengths with very low or null response level of the CCD.

\begin{figure*}[!htbp]
   \centering
\includegraphics[width=0.4\textwidth]{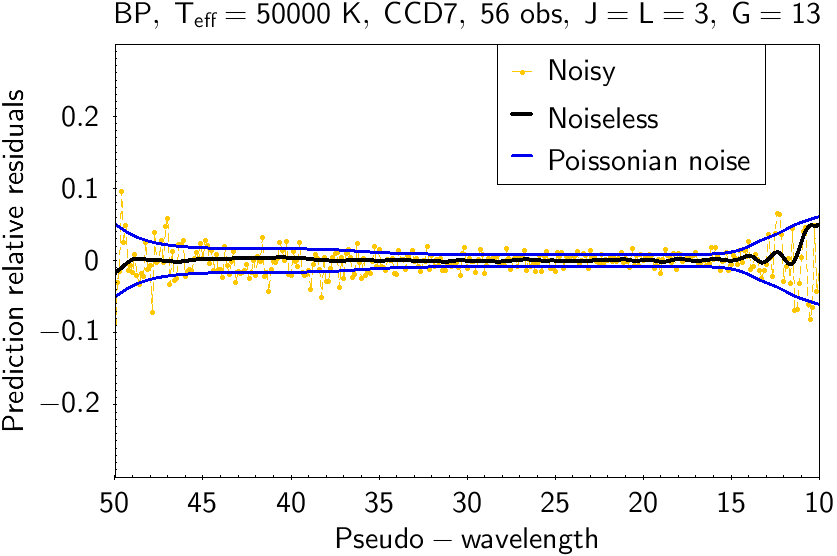}
\includegraphics[width=0.4\textwidth]{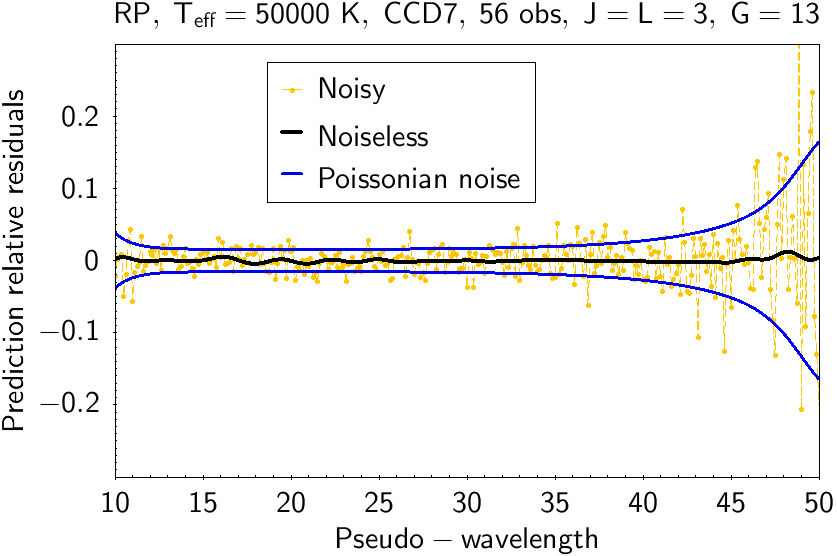}\\
\includegraphics[width=0.4\textwidth]{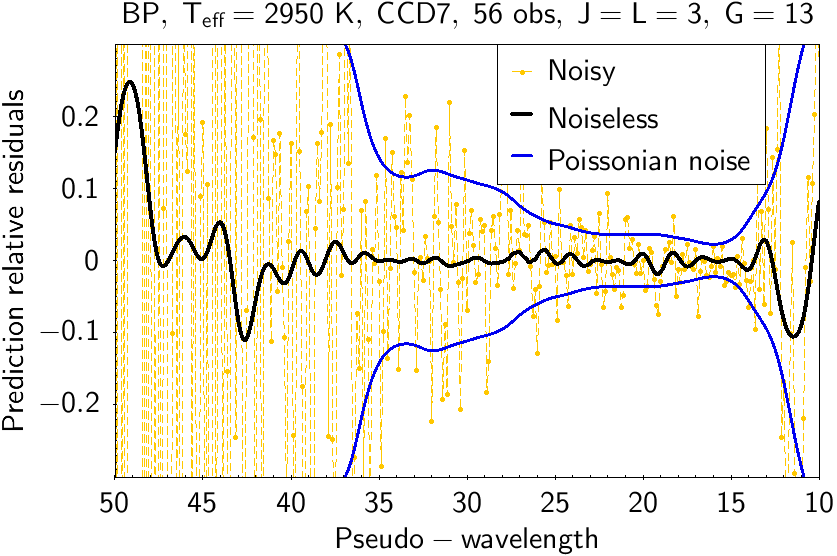}
\includegraphics[width=0.4\textwidth]{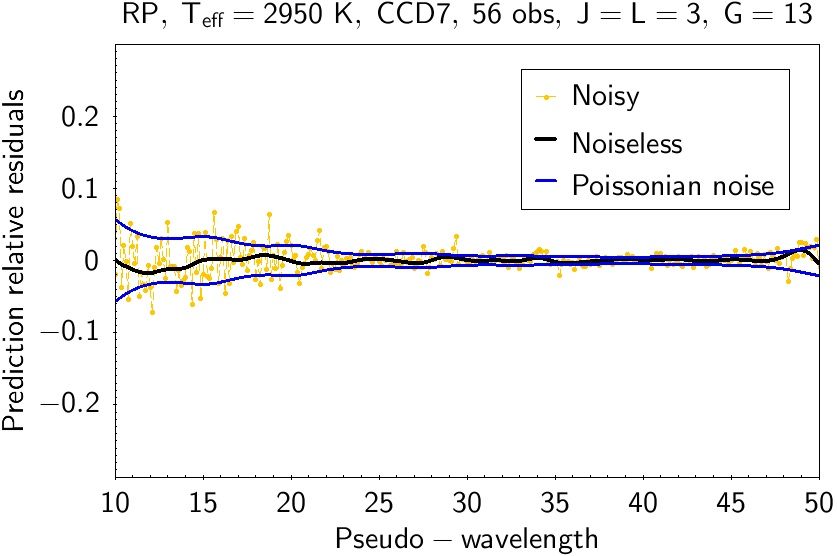}\\
\caption{Relative residuals per sample obtained for the predicted epoch observation in CCD7 for a hot (top) and a cold (bottom) $G=13$ mag simulated star using BaSeL-3.1 spectral energy distribution library \citep{Lejeune1997} for BP (left) and RP (right). $J$ and $L$ values refer to parameters in Eq.~(\ref{eq:basicConvolutionFinal}).
Pseudo-wavelength values smaller than 15 fall outside the nominal range of the BP instrument (having $\lambda>680$~nm). An approximate correspondence between pseudo-wavelength values and absolute wavelengths can be found in Fig.~\ref{fig:wavelength}.
  \label{fig:respredG13}
  }
\end{figure*}

\begin{figure*}[!htbp]
   \centering
\includegraphics[width=0.4\textwidth]{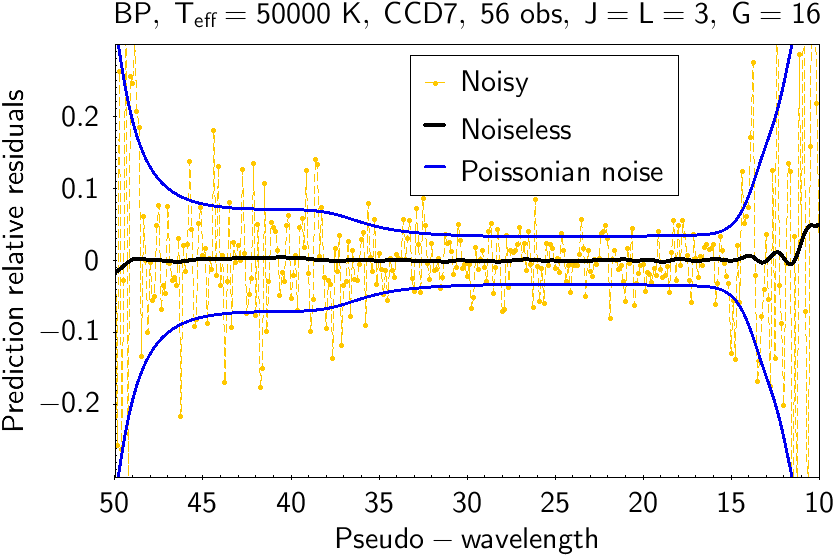}
\includegraphics[width=0.4\textwidth]{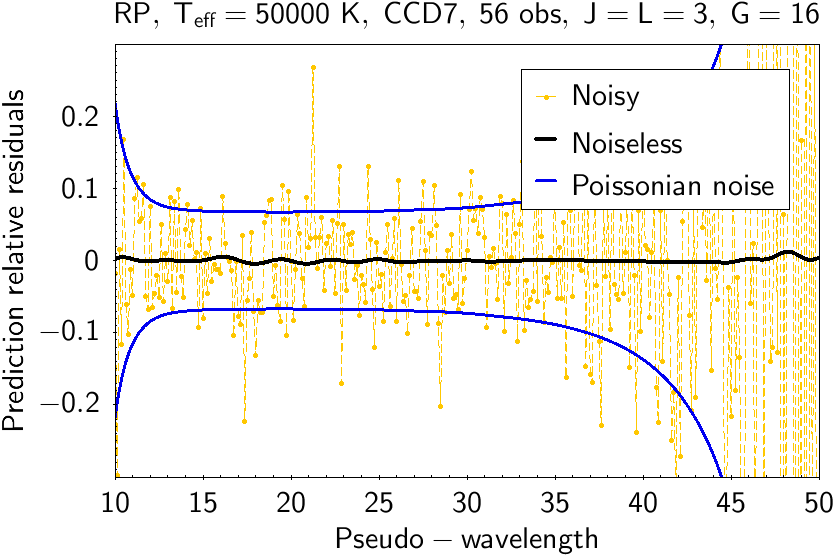}\\
\includegraphics[width=0.4\textwidth]{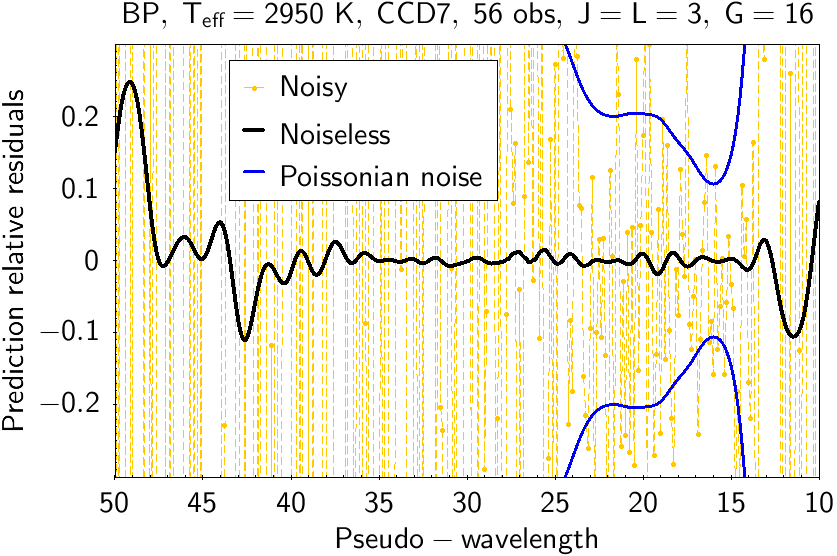}
\includegraphics[width=0.4\textwidth]{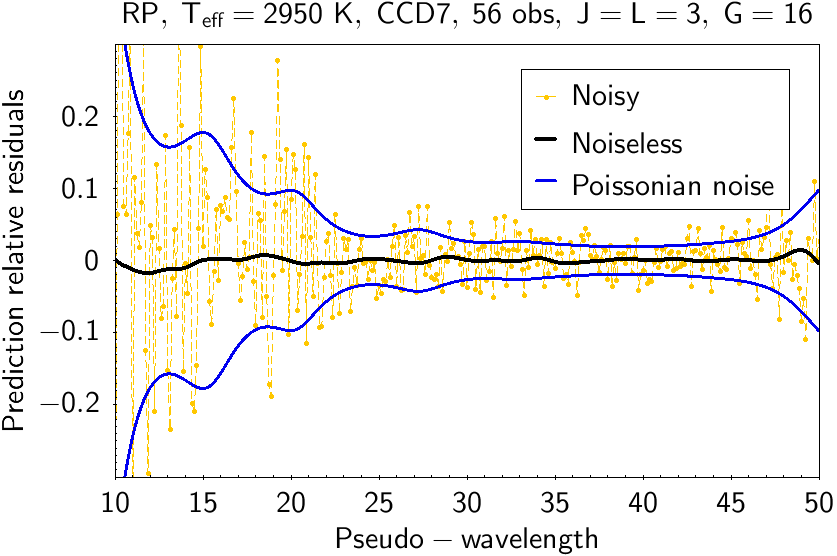}\\
\caption{Same as in Fig.~\ref{fig:respredG13}, but for $G=16$~mag.
  \label{fig:respredG16}
  }
\end{figure*}

\begin{figure*}[!htbp]
\centering
\includegraphics[width=0.32\textwidth]{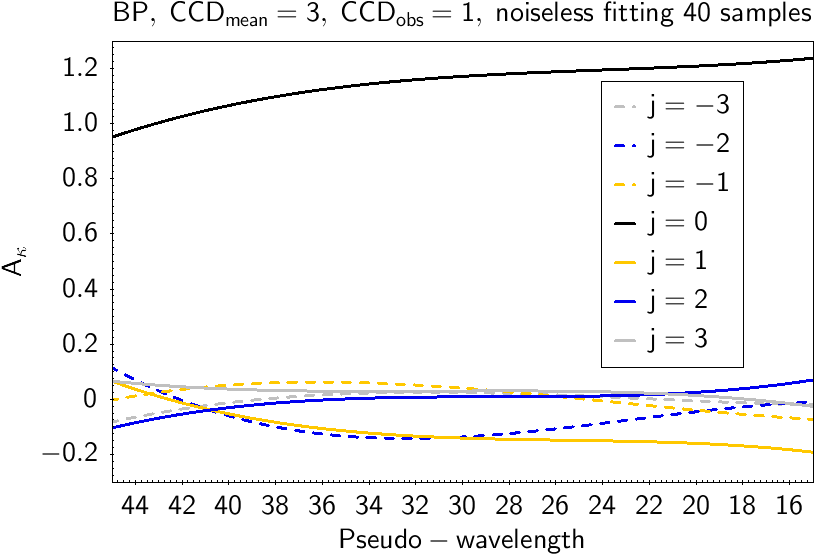}
\includegraphics[width=0.32\textwidth]{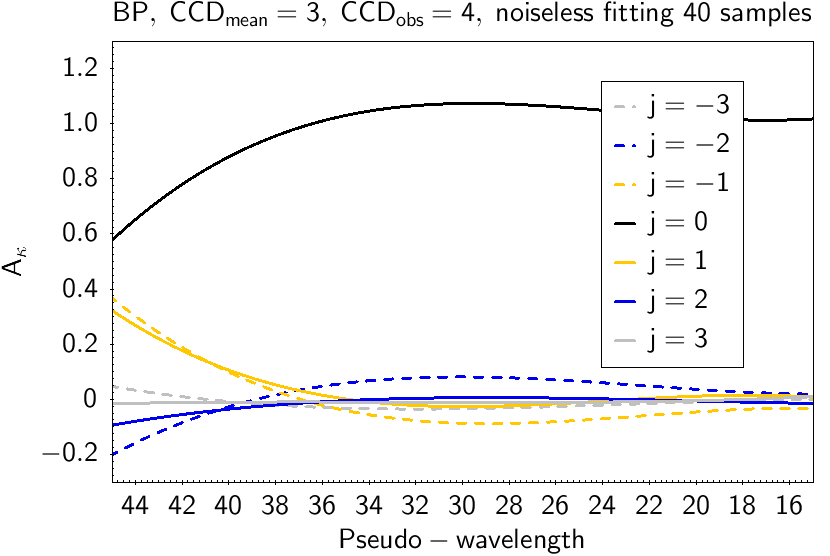}
\includegraphics[width=0.32\textwidth]{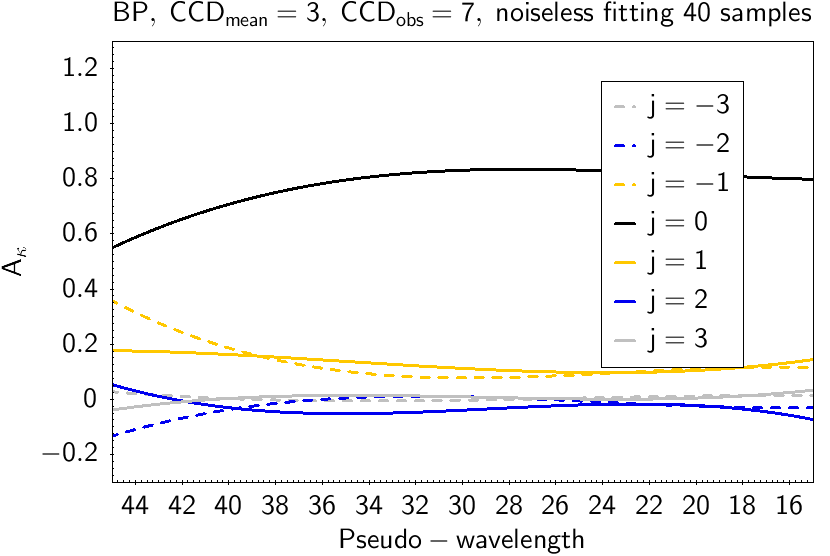}
	\caption{Instrumental $A_{\kappa}$ coefficients computed from simulations for 
BP CCD1 (left), CCD4 (centre), and CCD7 (right). 
\label{fig:aijresult}    
}
\end{figure*}

The instrumental and source coefficient updates have been iterated 10 times, although results already quickly converge to a stable solution for the second iteration.  
The $A_{\kappa}$ coefficients obtained after these iterations can be seen in Fig.~\ref{fig:aijresult}. The central sample evidently contributes the most, the neighbouring $\pm 1$ positions are the second most contributing positions, and so on, with contributions becoming smaller further away from the central sample. Extreme pseudo-wavelengths also deviate more from the nominal.

Figures~\ref{fig:respredG13} and \ref{fig:respredG16} show the relative residuals obtained when comparing both sides of Eq.~(\ref{eq:basicConvolutionFinal}) 
for two stars with the most different effective temperatures in the simulated dataset ($T_{\rm eff}=50\ 000$~K and $T_{\rm eff}=2950$~K) in CCD7 in the BP and RP spectrophotometers. 
The figures show the residuals for noiseless (black lines) and noisy (orange lines) observations corresponding to $G=13$~mag (Fig.~\ref{fig:respredG13}) and $G=16$~mag (Fig.~\ref{fig:respredG16}). 
The plot only shows the central pseudo-wavelength interval (from 10 to 50), as this is the region where most of the flux is located (see Fig.~\ref{fig:SpTspectra}) due to the passband response of the spectrophotometers. In fact, BP pseudo-wavelengths smaller than 15 go beyond the nominal response of the instrument ($\lambda>680$~nm), with pseudo-wavelength $u=10$ corresponding to a very large unrealistic wavelength ($\lambda\sim 870$~nm; as can be seen in Fig.~\ref{fig:wavelength}).

As expected, the residuals are large when there is little flux available (see, for example, the BP spectrum for a cold red source, $T_{\rm eff}=2950$~K, and the red RP spectrum for a hot blue source, $T_{\rm eff}=50\ 000$~K, and the wings of the spectra). When comparing the residuals with noise (in orange) with the Poissonian level\footnote{Although the simulations incorporate other components of the noise, as the background and read out noise (see \citealt{luri2014}), in Figs.~\ref{fig:respredG13} and ~\ref{fig:respredG16} (blue lines) we only compare with the Poissonian contribution as it directly 
relates with the amount of flux available.} (in blue), we see that the two quantities are compatible. The systematics present for the noiseless case (black lines) can only be improved if adding calibration sources with high signal-to-noise in the spectral regions with poor performance. This can be difficult to obtain; for instance, there are very few non-variable red sources in the Galaxy with enough flux in the blue to be used as calibrators and that introduce enough weight in the solution to represent this kind of sources.

\begin{figure*}[!htbp]
   \centering
\includegraphics[width=0.45\textwidth]{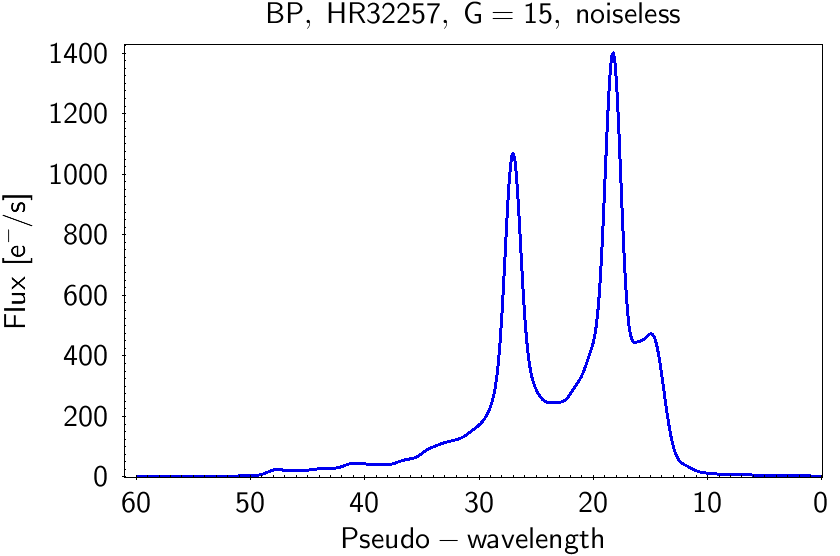}
\includegraphics[width=0.45\textwidth]{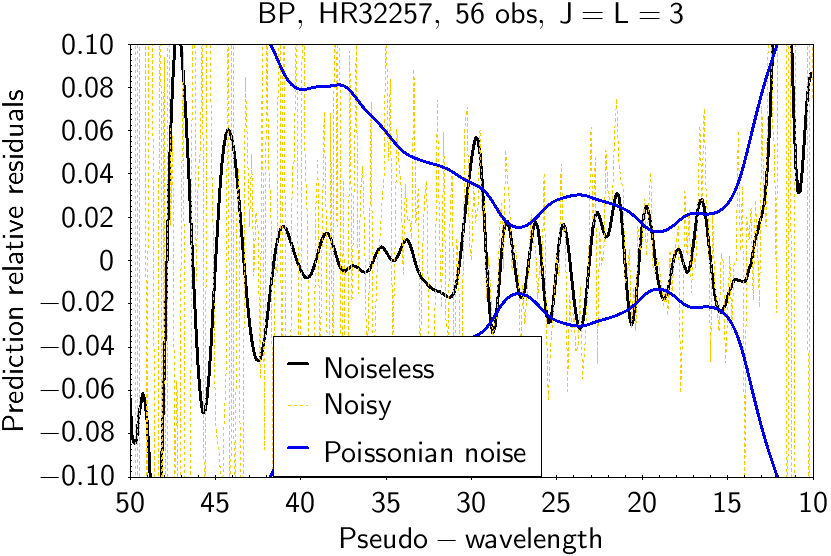}
\caption{Left: BP simulated spectrum with GOG \citep{luri2014} for a WC Wolf-Rayet star (HD32257). Right: Relative residuals obtained for the predicted epoch observation in CCD number 7.
We note that, in BP, the wavelength decreases when the pseudo-wavelength increases, and values smaller than 15 fall outside the nominal range of the instrument (with $\lambda>680$~nm), as can be seen in Fig.~\ref{fig:wavelength}.
  }
  \label{fig:respredWC4}
\end{figure*}

In Sect.~\ref{sec:calibrators}, we mention the importance of including all possible spectral shapes of the sources to be calibrated as calibrators, otherwise the residuals for those sources with shapes very different to the set of internal calibrators will include systematics. In order to illustrate this, we derived the residuals (see Fig.~\ref{fig:respredWC4}, right) for a Wolf-Rayet star (WC4=HD32257) simulated using GOG \citep{luri2014}. WC Wolf-Rayet stars have a spectrum dominated by emission lines of highly ionised elements: mainly He, C, and O. These features can be seen as strong features in the BP interval (see Fig.~\ref{fig:respredWC4}, left). 
The residuals show strong systematics of about 3--4\% in the emission lines wavelength range.

\subsection{Tests using {\Gaia} data}
\label{sec:testdata}
\begin{figure*}[!htbp]
   \centering
\includegraphics[width=0.49\linewidth]{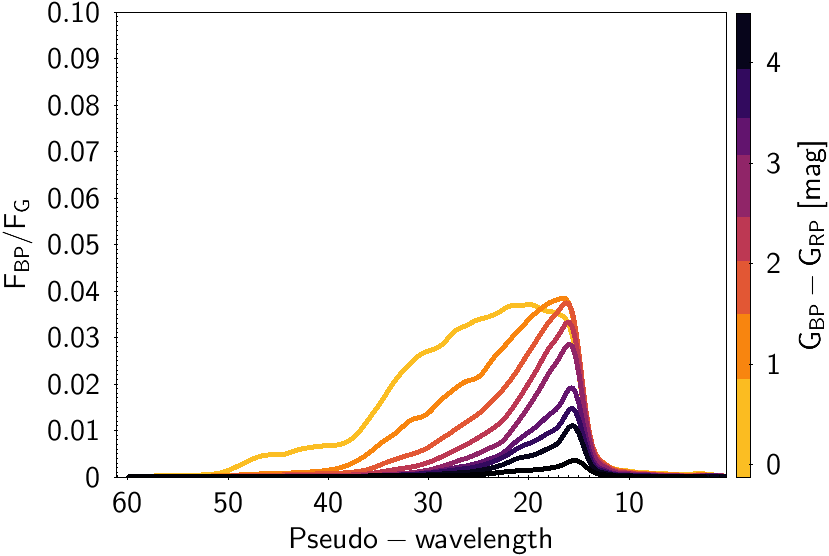}
\includegraphics[width=0.49\linewidth]{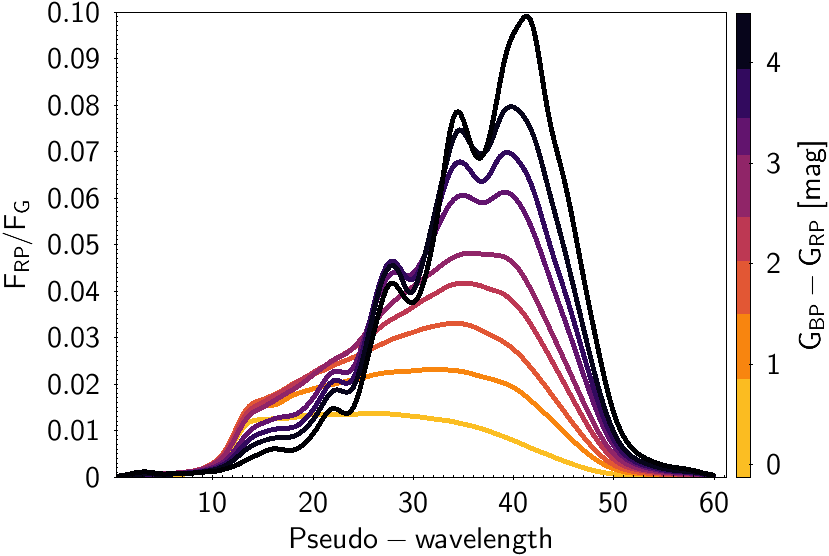}
\caption{Preliminary mean spectra obtained from {\Gaia} observations, normalised to the flux in the $G$ passband. 
This data was extracted from cycle 3, 
processing for sources with different $G_{\rm BP}-G_{\rm RP}$ colours (see colour bar) for BP (left) and RP (right). An approximate correspondence between pseudo-wavelength values and absolute wavelengths can be found in Fig.~\ref{fig:wavelength}.
  }
  \label{fig:SpTspectra}
\end{figure*}

Figure~\ref{fig:SpTspectra} shows mean spectra obtained for some actual {\Gaia} spectrophotometric observations to be published in {\GDR3}. The figure shows the mean spectra obtained for sources with different colours, 
normalised by the total flux in $G$. All sources in Fig.~\ref{fig:SpTspectra} have a magnitude close to $G=13$ mag. 
In the absence of extinction, the colour of the stars is related with their effective temperatures, with the hot stars being
bluer than cooler ones. Thus, hot stars have most of their flux in BP; cooler 
ones, on the contrary, are brighter in RP. This is a gradual effect and intermediate temperatures have 
a more balanced contribution in both wavelength ranges. 
The expected signal-to-noise ratio in the {\GDR3} mean spectra close to the peak of the flux distribution at this magnitude is about 600 for BP and 1000 for RP.

\begin{figure*}[!htbp]
   \centering
\includegraphics[width=0.49\linewidth]{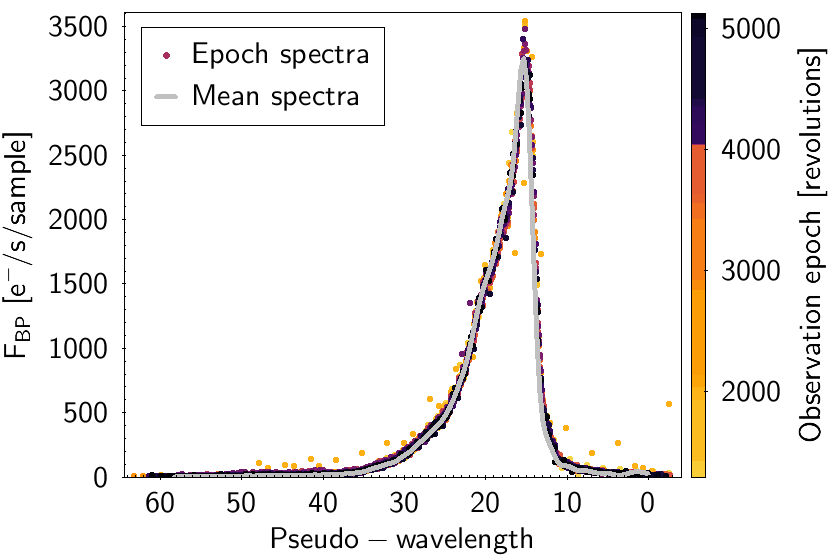}
\includegraphics[width=0.49\linewidth]{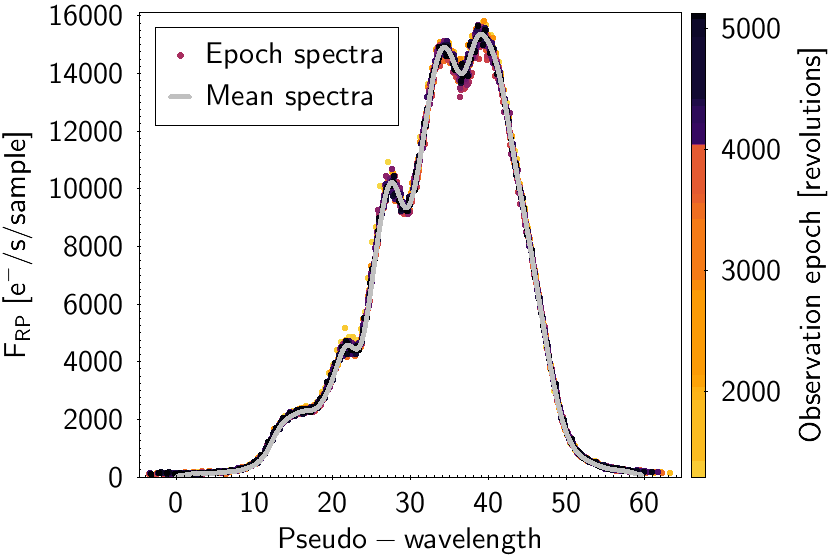}
\caption{Calibrated epoch (dots) and mean spectra (grey line) for the source with $G_{\rm BP}-G_{\rm RP}=3.5$~mag also plotted in Fig.~\ref{fig:SpTspectra}. The different epochs in {\Gaia} are measured 
in a number of six-hour revolutions done by the satellite while surveying the sky. An approximate correspondence between pseudo-wavelength values and absolute wavelengths can be found in Fig.~\ref{fig:wavelength}.
  }
  \label{fig:MSspectra}
\end{figure*}

Figure~\ref{fig:MSspectra} shows an example to demonstrate how these mean spectra are derived from the epoch observations. The figure shows the epoch spectra (dots) used to derive the mean spectra (grey line) for the source in Fig.~\ref{fig:SpTspectra} with $G_{\rm BP}-G_{\rm RP}=3.5$~mag. The obtained mean spectra represent the shape of the spectra in the mean instrument.

\begin{figure*}[!htbp]
   \centering
\includegraphics[width=0.49\linewidth]{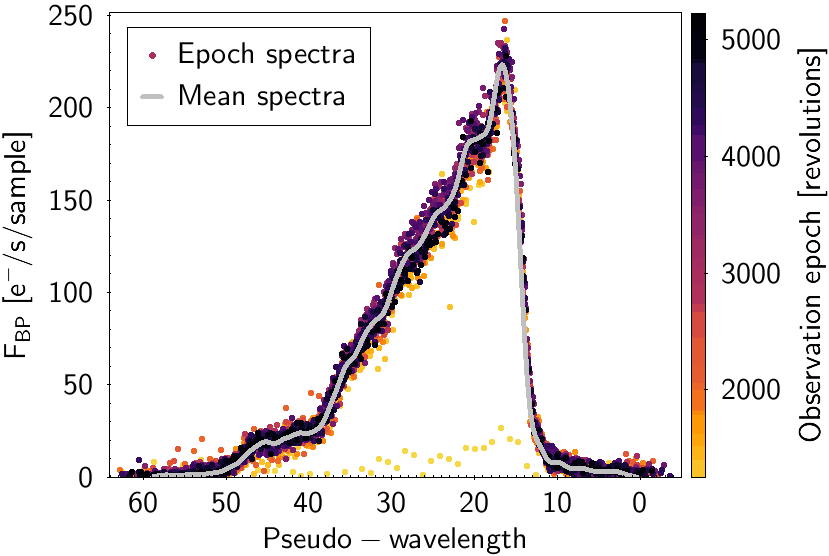}
\includegraphics[width=0.49\linewidth]{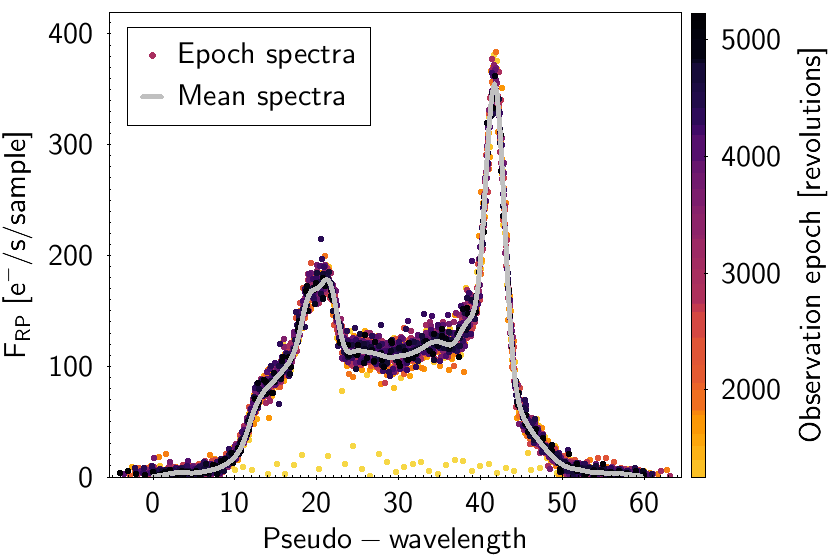}
\caption{Epoch and mean spectra for an extragalactic source from the {\Gaia} Celestial Reference Frame ({\Gaia}-CRF, \citealt{mignard18}).
The presence of an outlier observation with a very low flux level does not disturb the derivation of its mean spectra (see text). The different epochs in {\Gaia} are measured 
in number of six-hour revolutions done by the satellite while surveying the sky. An approximate correspondence between pseudo-wavelength values and absolute wavelengths can be found in Fig.~\ref{fig:wavelength}.
  \label{fig:QSOspectra}
  }
\end{figure*}

\begin{figure*}[!htbp]
   \centering
\includegraphics[width=0.49\linewidth]{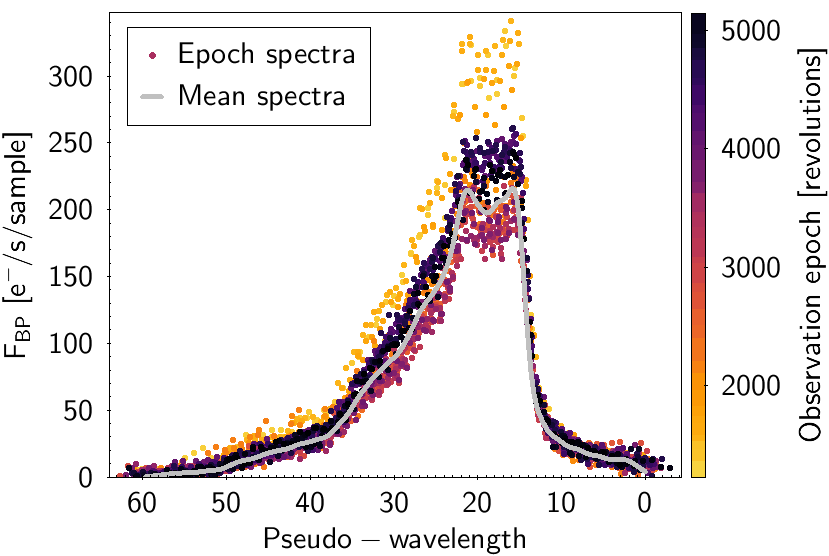}
\includegraphics[width=0.49\linewidth]{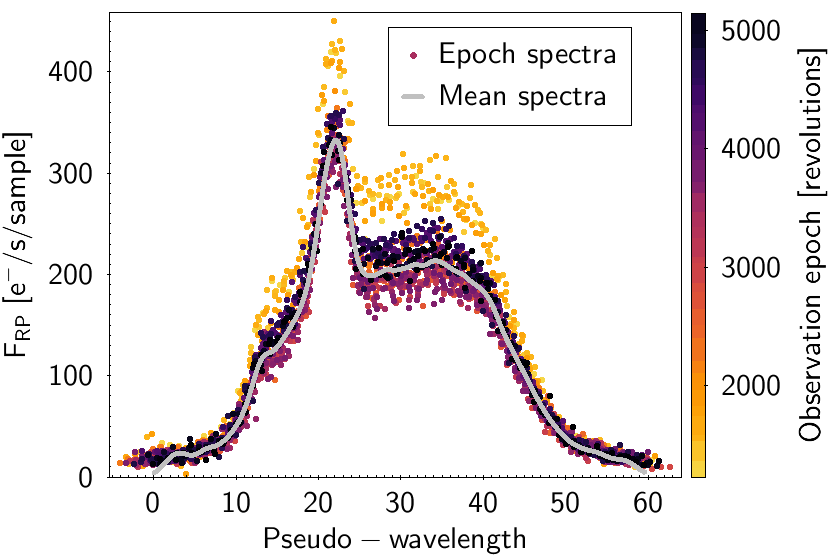}
\caption{Same as in Fig.~\ref{fig:QSOspectra}, but for a different quasar. 
The flux-level variations between observations made at different epochs indicate the 
variable nature of the source.
  }
  \label{fig:QSOspectraVariable}
\end{figure*}

{\Gaia} also observes point-like sources that are different to stars.
For example, extragalactic sources (far-away quasars) are used to establish 
the astrometric reference frame ({\Gaia} Celestial Reference Frame, {\Gaia}-CRF, \citealt{mignard18}). 
For those sources, the spectral energy distribution in BP and RP differs drastically 
from those of stellar sources. Some of these quasars can have strong emission lines and experience flux variability. 
Two of these quasars are plotted in Figs.~\ref{fig:QSOspectra} and \ref{fig:QSOspectraVariable}.
They show both mean and epoch spectra over-plotted, illustrating the result of building 
the mean spectra from different epoch observations. 
The source in Fig.~\ref{fig:QSOspectra} has a very low flux level in one of its observations at early stages of the mission.
The photometric processing is sufficiently robust to consider this measurement as an outlier, 
thereby not contributing to the mean spectrum. 
The source plotted in Fig.~\ref{fig:QSOspectraVariable} is a variable one changing its flux level at different epochs. 
Although the mean spectra may not be representative of any particular epoch, 
they are useful to provide a hint of the mean spectral shape and characteristics of the source.

\section{Summary and conclusions}
\label{sec:conclusions}

The model explained here deals with the internal calibration of the {\Gaia} BP/RP spectra. 
Its goal is to combine all epoch spectra obtained with very different conditions into a unique instrumental 
system that could later on be converted into an absolute system of fluxes and wavelengths 
by the external calibration. The details of this complex external calibration step will be provided in a future paper (Montegriffo et al, in preparation). Nevertheless, for some applications it could be more convenient to work directly with the internally calibrated spectra, was as done by the CU8 {\Gaia} team to derive the astrophysical parameters (for example, using the Apsis algorithm, \citealt{Andrae2018}).

The outputs of the internal calibration model explained in this paper are the instrumental coefficients at every calibration unit, which allow the computation of the mean spectra for every observed source.
The model consists of a forward model to predict every observation 
and minimise the residuals with actual observations. 
The two main ingredients in this forward model are the 
description of the source (using a set of basis functions) and the instrument model (which 
accounts for differential response, dispersion, LSF, AL, and AC flux loss, and small geometric shifts). These two pieces are iterated until convergence, using a set of millions of calibration sources covering a range of spectral energy distributions as wide as possible. 

The first BP- and RP-calibrated spectra using the model described in this paper will be published in {\GDR3} (expected for 2022). 
Preliminary results using data from the {\Gaia} mission confirm the suitability of this approach. It has also been confirmed that using the information from the mean spectra, the astrophysical information can be recovered.

The calibration technique outlined in this work differs significantly from commonly used approaches of spectral calibration, which is a consequence of the self-calibrating nature of the low-resolution {\Gaia} spectrophotometers. An important implication of this approach is the fact that the calibrated spectra are obtained as a linear combination of basis functions, rather than a table of wavelengths and fluxes. This avoids the loss of information produced when sampling the spectra from the continuous representation provided by the basis functions. The number of coefficients can be reduced using the optimisation strategy described in Sect.~\ref{sec:source}.  

Our recommendation is that users familiarise themselves with adopting continuous representation of the spectra, instead of the sampled spectra, as it contains the full range of information contained in the spectra, including the effect of the astrophysical parameters. Nevertheless, for those users for whom the loss of information while sampling the spectra is not critical, the {\Gaia} Data Processing and Analysis Consortium (DPAC) community will also provide tools\footnote{Together with {\GDR3}, DPAC will provide the scientific community with a set of tools to produce sampled spectra (both internally and externally calibrated) and to transform from internal to external and vice versa.} to derive sampled spectra from the coefficients.

\begin{acknowledgements}
This work was done inside the {\Gaia} Data Processing and Analysis Consortium (DPAC) and the authors want to thank all members in that collaboration for the fruitful discussions done in several parts of the process until converging to the model presented in this paper. In particular we want to thank all members in CU5, CU8 and CU9 units and the Editorial Board in DPAC providing feedback on this work.
We want to thank especially Ren\'e Andrae, Rosanna Sordo, Morgan Fouesneau, Ludovic Delchambre, Yves
Fremat, Luis Manuel Sarro, Alex Lobel and Coryn A.L.~Bailer-Jones who helped testing the different possible configurations of the model proposed here for its implementation in {\GDR3}.
      Authors at the ICCUB were supported by the Spanish Ministry of Science, Innovation and University (MICIU/FEDER, UE) through grant RTI2018-095076-B-C21, and the Institute of Cosmos Sciences University of Barcelona (ICCUB, Unidad de Excelencia ’Mar\'{\i}a de Maeztu’) through grant CEX2019-000918-M.      
      Authors at the Institute of Astronomy of the University of Cambridge were supported by the United Kingdom Science and Technology Facilities Council (STFC) and the
United Kingdom Space Agency (UKSA) through the following grants to the
University of Cambridge: ST/K000756/1, ST/N000641/1 and ST/S000089/1.
Authors at INAF were supported by the Agenzia Spaziale Italiana (ASI) through contracts I/037/08/0, I/058/10/0, 2014-025-R.0,
2014-025-R.1.2015  and 2018-24-HH.0 to the Italian Istituto Nazionale di Astrofisica (INAF), 
and INAF.
\end{acknowledgements}

\bibliographystyle{aa} 
\bibliography{bibliography} 
%
%


\begin{appendix}

\section{Nomenclature}
Tables~\ref{tab:acronyms} and \ref{tab:variables} include a summary of the list of acronyms and variables used in this paper.

\begin{table}
\begin{center}
\caption{Meanings of the several acronyms used in this paper.
\label{tab:acronyms}}
\begin{tabular}{ll}
\hline
Acronym    & Definition\\
\hline
AC  & Across scan direction\\
AL  & Along scan direction\\
BP  & Blue photometer \\
CCD & Charged coupled device\\
CRNU & Column response non uniformity \\
DR & Data release\\
EDR & Early data release\\
ESA & European space agency\\
FoV & Field of view\\
GOG & {\Gaia} Object Generator\\
LSF & Line spread function\\
PSF & Point spread function\\
RP  & Red photometer \\
SPD & Spectral photon distribution\\
SVD & Singular value decomposition\\
TDI & Time delay integration\\
\hline          
\end{tabular}
\end{center}
\end{table}

\begin{table*}
\begin{center}
\caption{Meaning of the several parameters, variables, and indices used in this paper. 
\label{tab:variables}}
\begin{tabular}{ll}
\hline
Variable  & Definition\\
\hline
$\alpha$ & Index to describe the colour dependence of the flux loss\\
$\beta$ & Index to describe the along scan position dependence of the along scan flux loss\\
$\gamma$ & Index to describe the AC centring error dependence of the AC flux loss\\
$\Delta$ & Total bandwidth in wavelength units\\
$\Delta x$ & Across scan centring error\\
$\Delta y$ & Along scan centring error\\
$\Delta \theta$ & Shift to transform pseudo-wavelength scale to the independent variable to describe Hermite functions\\
$\delta$ & Kronecker delta function\\
$\Phi$ & $N\times1$ matrix with all transformed source basis functions to a given calibration unit\\
$\varphi$ & Source basis function\\
$\bar{\varphi}$ & Source basis function transformed to a given calibration unit\\
$\kappa$ & Calibration Unit\\
$\Lambda$ & Line Spread Function\\
$\lambda$  & Wavelength\\
$\lambda_{\rm cuton}$  & Cut-on wavelength of the {\Gaia} passbands\\
$\lambda_{\rm cutoff}$  & Cut-off wavelength of the {\Gaia} passbands\\
$\mu$ & Mean spectra scale\\
$\nu$ & Index to describe the AC velocity dependence of the AC flux loss\\
$\omega$ & Each of the parameters needed to describe the instrument\\
$\Omega$ & Number of parameters needed to describe the instrument\\
$\sigma$ & Standard deviation\\
$\theta$ & Independent variable of the Hermite functions\\
$\Theta$ & Scaling parameter\\
$A$ & Kernel to transform flux from mean scale to a given calibration unit\\
$B$ & $M\times N$ matrix with all source coefficients for all internal calibrators\\
$\bf C$ & Hermite coefficients matrix \\
$D$ & Relative dispersion with respect to nominal\\
$E$ & Functions to describe the dependence of $A$ with pseudo-wavelength\\
$G$ & {\Gaia} white light magnitude\\
\BP & Integrated blue Photometer magnitude\\
\RP & Integrated red Photometer magnitude\\
$H$ & $M \times 1$ matrix with epoch spectra in a given calibration unit for all internal calibrators\\
$I$ & Instrument kernel function\\
$J$ & Maximum number of neighbours at each side of the pseudo-wavelength to be predicted\\
$K+1$ & Number of coefficients to describe the across dependence of the instrument\\
$L+1$ & Total number of $B$ functions\\
$M$ & Number of internal calibration sources\\
$N+1$ & Total number of source basis functions\\
$N_{\alpha}$ & Degree of the AL flux loss polynomial dependence with colour\\
$N'_{\alpha}$ & Degree of the AC flux loss polynomial dependence with colour\\
$N_{\beta}$ & Degree of the AL flux loss polynomial dependence with AL centring error\\
$N_{\gamma}$ & Degree of the AC flux loss polynomial dependence with AC centring error\\
$N_{\nu}$ & Degree of the AC flux loss polynomial dependence with AC velocity\\
$N_{\rm src}$ & Number of sources\\
$P$ & Along scan flux loss\\
$Q$ & Across scan flux loss\\
$R$ & Response function\\
$S$ & Spectral Photon Distribution\\
$T_{\rm eff}$ & Effective temperature\\
$U$ & Number of pseudo-wavelengths sampling points\\
$\bf V_C$ & Rotated matrix of the Hermite coefficients matrix \\
$a$ & Allowed threshold for the standard deviation of $r$ when truncating the source coefficients\\
$b$ & Source coefficients\\
$c$ & Coefficient to describe the dependence with pseudo-wavelength of the instrument\\
$d$ & Coefficient to describe the across scan dependence of the instrument\\
$h$ & Epoch spectrum\\
$i$ & Index for the pixel \\
$j$ & Index for the neighbour sample\\
$k$ & Index for the across scan dependence of the instrument\\
$l$ & Index for the $B$ basis functions\\
$m$ & Index for the internal calibration source \\
$n$ & Index for the source basis function\\
$n^{\prime}$ & Number of leading source basis functions selected during the truncation procedure\\
$p$ & Coefficient to describe the along scan flux loss\\
$q$ & Coefficient to describe the across scan flux loss\\
$r$ & Normalised standard deviation of the source coefficients excluding basis functions\\
$s$ & Index for the source.\\
$t$ & Index for infinite number of neighbours\\
$u$ & Pseudo-wavelength\\
$v_{AC}$ & Across scan velocity\\
$x$ & Across scan position\\
\hline          
\end{tabular}
\end{center}
\end{table*}

\end{appendix}

\end{document}